\documentclass[aps,pra,reprint,superscriptaddress,amsmath,amssymb]{revtex4-2}

\usepackage[english]{babel}

\usepackage{mathrsfs}
\usepackage{amsfonts}
\usepackage{amsmath}
\usepackage{amsthm}
\usepackage{amssymb}
\usepackage{dsfont}
\usepackage{esint}
\usepackage{bbm}
\usepackage{bm}
\usepackage{mathtools}

\usepackage{enumerate}
\usepackage[shortlabels]{enumitem}
\usepackage{framed}
\usepackage{csquotes}

\usepackage{tabularx}
\usepackage[dvipsnames]{xcolor}
\usepackage{hyperref}
\definecolor{links}{rgb}{0.36,0.54,0.66}
\hypersetup{
   colorlinks = true,
    linkcolor = purple,
     urlcolor = blue,
    citecolor = blue,
    filecolor = blue,
    pdfauthor = {Author},
     pdftitle = {Title},
   pdfsubject = {subject},
  pdfkeywords = {one, two},
  pdfproducer = {LaTeX},
   pdfcreator = {pdfLaTeX},
   }
\newtheorem{theorem}{Theorem}

\newtheorem{lemma}[theorem]{Lemma}

\newtheorem{proposition}[theorem]{Proposition}

\newcommand{\pd}{\partial}
\newcommand{\RR}{\mathbb{R}}
 
\newcommand{\NN}{\mathbb{N}} 
 
\newcommand{\ii}{\mathbbm{1}}

\newcommand{\op}[2]{|#1\rangle\langle #2|} 
\newcommand{\ip}[2]{\left\langle #1\middle| #2 \right\rangle} 
 
\newcommand{\Tr}{\operatorname{Tr}}

\newcommand{\abs}[1]{\left|#1\right|} 
\newcommand{\ra}{\rightarrow}

\newcommand{\ph}{\varphi}
\newcommand{\bp}[1]{\!\left(#1\right)} 
\newcommand{\bb}[1]{\!\left[#1\right]}
\newcommand{\bc}[1]{\!\left\{#1\right\}} 
\newcommand{\ba}[1]{\!\left\langle #1\right\rangle}

\allowdisplaybreaks

\begin{document}

\title{Quantum Fisher information matrices for\\time-evolved bosonic Gaussian thermal states}

\author{Hailey S.~Murray}
\affiliation{School of Applied and Engineering Physics, Cornell University, Ithaca, NY 14853, USA}
\author{Diego Estrada}
\affiliation{Department of Physics, University of Maryland, Baltimore County, Baltimore, Maryland 21250, USA}
\affiliation{College of Arts and Sciences, Cornell University, Ithaca, New York 14853, USA}
\author{Mark M.~Wilde}
\affiliation{School of Electrical and Computer Engineering,
Cornell University, Ithaca, New York 14853, USA}

\date{\today}

\begin{abstract}
    Quantum Fisher information (QFI) is a fundamental quantum metrological quantity arising from the second order term in a Taylor expansion of a smooth divergence evaluated at nearby quantum states.
    The $\alpha$-$z$ information matrix is a QFI measure arising from the $\alpha$-$z$ R\'enyi relative entropy, and QFI measures corresponding to various underlying smooth divergences may be recovered from the $\alpha$-$z$ information matrix, including the standard symmetric logarithmic derivative (SLD) QFI. Bosonic Gaussian states are common probe states for quantum metrological tasks and admit a complete characterization in terms of their first and second moments. In this work, we consider the quantum metrological task of estimating parameters that characterize a Gaussian unitary time evolution of a faithful bosonic Gaussian thermal state. We establish a formula for the $\alpha$-$z$ information matrix of faithful bosonic Gaussian thermal states that have undergone a Gaussian unitary time evolution. As an example, we calculate entries of the $\alpha$-$z$ information matrix for a two-mode Gaussian thermal state that has undergone a beam splitting operation. For a parametrized family of time-evolved bosonic Gaussian states, we also derive a formula for the SLD operator, which is known to provide a measurement scheme that saturates the quantum Cr\'amer--Rao bound and thus achieves the ultimate precision for single parameter estimation.
\end{abstract}

\maketitle

\section{Introduction}

\subsection{Background}

Fisher information measures the sensitivity of a probability distribution to changes in unknown parameters upon which the distribution depends. It is useful in many contexts, including parameter estimation, machine learning, and optimization~\cite{korostelev2011,amari1998}. Classically, the Fisher information arises from the second order term in a Taylor expansion of a  smooth divergence evaluated at nearby probability distributions~\cite{cencov1978}. The quantum analogue of the Fisher information, known as the quantum Fisher information (QFI), measures the sensitivity of a quantum state to changes in unknown parameters upon which the state depends (see, e.g.,~\cite{sidhu2020} for a review). It arises from the second order term in a Taylor expansion of a smooth divergence evaluated at nearby quantum states. Because classical divergences have no unique quantum generalization, the QFI is not unique~\cite{petz1996}. The standard symmetric logarithmic derivative (SLD) QFI arises from the Bures distance, which is a distinguishability measure for quantum states that can be expressed in terms of the fidelity. When the underlying smooth divergence is instead the $\alpha$-$z$ R\'enyi relative entropy, the resulting QFI measure is known as the $\alpha$-$z$ information matrix~\cite{wilde2025}. Specific choices of $\alpha$ and $z$ recover the SLD QFI as well as the QFI matrices corresponding to the Petz-- and sandwiched R\'enyi relative entropies~\cite{wilde2025}.

The Cr\'amer--Rao bound sets a limit on how well one can estimate an unknown parameter upon which a probability distribution depends. Specifically, the Cr\'amer--Rao bound states that the variance of an unbiased estimator of an unknown parameter is bounded from below in terms of the inverse Fisher information. Similarly, the quantum Cr\'amer--Rao bound sets the limit on how well one can estimate an unknown parameter using a quantum state that depends on the parameter. The quantum Cr\'amer--Rao bound states that the variance of an unbiased estimator of an unknown parameter is bounded from below in terms of the inverse of the SLD QFI~\cite{helstrom1967,helstrom1968}. For the estimation of a single parameter, it is known that the SLD operator saturates the quantum Cr\'amer--Rao bound~\cite{braunstein_caves1994}. In general, however, the SLD operator may depend on the unknown parameter. 

Bosonic Gaussian states are commonly used as probe states for quantum metrological tasks~\cite{safranek2015quantumparameterestimation,oh2019}, and they constitute a family of states that are of fundamental importance for experimental and theoretical quantum information science~\cite{wang_hiroshima_tomita_hayashi2007,adesso_ragy_lee2014,serafini2017,weedbrook_pirandola2012}. Because they are completely characterized by their first and second moments, calculations involving bosonic Gaussian states are more mathematically tractable than the general non-Gaussian case. Continuous-variable bosonic systems are present in many quantum information processing architectures, including optical modes \cite{CMP14,Yokoyama2013,Roslund2014}, vibrational modes of trapped ions \cite{Fluehmann2019}, microwave-cavity modes \cite{Rosenblum2018}, microwave resonators coupled to superconducting Josephson junctions \cite{Wilson2011,GU20171}, and opto-mechanical resonators \cite{AKM14,HAF15}.

\subsection{Summary of results}
The metrological task that we consider is the estimation of parameters that govern a Gaussian unitary evolution of a faithful bosonic Gaussian thermal state. Examples of such an evolution include the estimation of unknown parameters in an interferometer, coupling rates between modes, or squeezing parameters. Due to the breadth of this framework, having a general expression for QFI would be helpful in many quantum metrological scenarios. In this work, we establish a formula for the entries of the $\alpha$-$z$ Fisher information matrix for bosonic Gaussian thermal states that have undergone a Gaussian unitary evolution (Theorem~\ref{thm:main}). We also calculate the SLD operator for time-evolved bosonic Gaussian thermal states (Proposition~\ref{prop:sld}). As an example, we calculate entries of the $\alpha$-$z$ information matrix for a two-mode thermal state undergoing a beam splitting operation.

\subsection{Paper organization}
Our paper is organized as follows. In Section~\ref{sec:prelim}, we cover preliminary topics. In Section~\ref{sec:res}, we provide our main results, which are formulas for the $\alpha$-$z$ information matrix and SLD operator for time-evolved bosonic Gaussian thermal states, and we include an example of calculating the $\alpha$-$z$ information matrix entries for a two-mode thermal state passing through a beam splitter. We then conclude with final comments and suggestions for future research directions in Section~\ref{sec:concl}.

\section{Preliminaries}\label{sec:prelim}

\subsection{Bosonic Gaussian states}\label{subsec:prelim_bos_gauss}
We will begin by reviewing some preliminary facts about bosonic Gaussian states. For an $n$-mode bosonic Gaussian system, where $n\in\NN$, we let $\hat{r}$ denote the vector of canonical position and momentum quadrature operators:
\begin{align}
    \hat{r}\coloneq\bb{\hat{x}_1,\hat{p}_1,...,\hat{x}_n,\hat{p}_n}^T,
\end{align}
where $\hat{x}_i$ and $\hat{p}_i$ are the position and momentum quadrature operators acting upon mode $i$, respectively. These operators satisfy the canonical commutation relations:
\begin{align}
    \bb{\hat{r}_i,\hat{r}_j}&=i\Omega_{i,j}.
\end{align}
Here, $\Omega$ is the symplectic form given by
\begin{align}
    \Omega&\coloneqq \bigoplus_{k=1}^n\Omega_1\\
    &=\ii_n\otimes\Omega_1,
\end{align}
and
\begin{align}
    \Omega_1&\coloneq\begin{bmatrix}
        0 & 1 \\ -1 & 0
    \end{bmatrix}.
\end{align}
An $n$-mode bosonic Gaussian state $\rho$ is fully characterized by its $2n\times1$ mean vector and $2n\times2n$ covariance matrix. The entries of the mean vector $\mu$ and covariance matrix $V$ of $\rho$ are defined as
\begin{align}
    \mu_i&\coloneq\ba{\hat{r}_i}_\rho,\\
    V_{j,k}&\coloneq\frac{1}{2}\ba{\,\bc{\hat{r}_j^{c_\mu},\hat{r}_k^{c_\mu}}}_\rho,
\end{align}
where we define
\begin{align}
    \hat{r}^{c_\mu}&\coloneq \hat{r}-\mu.\label{def:centered_vec}
\end{align}
Every bosonic Gaussian state $\rho$ with covariance matrix $V$ satisfies the following uncertainty relation:
\begin{align}
    V+\frac{i}{2}\Omega\geq0.
\end{align}
If $\rho$ is faithful, then the inequality above is strict.

In general, an $n$-mode faithful bosonic Gaussian thermal state $\rho$ with mean vector $\mu$, covariance matrix $V$, and real, symmetric, and positive definite Hamiltonian matrix $G$ has the form
\begin{align}
    \rho(\mu,G)&\coloneq\frac{1}{Z(\mu,G)}\exp\bp{-\frac{1}{2}\bp{\hat{r}^{c_\mu}}^TG\hat{r}^{c_\mu}},
    \label{def:rho_faithful_gaussian}
\end{align}
where $Z(\mu,G)$ denotes the partition function of $\rho$, which is defined as
\begin{align}
    Z(\mu,G)\coloneq\sqrt{\det\bp{V+\frac{i\Omega}{2}}}.\label{def:partiton}
\end{align}
The Hamiltonian matrix and covariance matrix of $\rho$ satisfy the following relationships: 
\begin{align}
    G&=2i\Omega\operatorname{arccoth}(2iV\Omega),\\
    V&=\coth\bp{\frac{i\Omega G}{2}}\frac{i\Omega }{2}.
\end{align}

\subsection{\boldmath$\alpha$-$z$ information matrix}

The $\alpha$-$z$ R\'enyi relative entropy is a quantum generalization of the classical R\'enyi divergence. Let $\alpha\in(0,1)\cup(1,\infty)$ and $z>0$. For all positive definite quantum states $\rho$ and $\sigma$, the $\alpha$-$z$ R\'enyi relative entropy is defined as~\cite{audenaert_datta2015}
\begin{align}
    D_{\alpha,z}(\rho\|\sigma)&\coloneq\frac{1}{\alpha-1}\ln\Tr\bb{\bp{\sigma^{\frac{1-\alpha}{2z}}\rho^{\frac{\alpha}{z}}\sigma^{\frac{1-\alpha}{2z}}}^z}.
\end{align}
The Petz--R\'enyi relative entropy~\cite{petz1985,petz1986} is recovered from the $\alpha$-$z$ R\'enyi relative entropy when $z=1$, and the sandwiched R\'enyi relative entropy~\cite{mullerlennert2013,wilde_winter_yang2014} is recovered when $z=\alpha$~\cite{audenaert_datta2015}.

For a family $\bp{\sigma(\phi)}_{\phi}$ of parametrized positive definite states, the entries of the $\alpha$-$z$ information matrix are defined as~\cite[Eq.~(5.4)]{wilde2025}
\begin{align}
    \bb{I_{\alpha,z}(\theta)}_{i,j}\coloneq \left. \frac{1}{\alpha}\frac{\pd^2}{\pd\varepsilon_i\pd\varepsilon_j}D_{\alpha,z}\bp{\rho(\theta)\middle\|\rho(\theta+\varepsilon)}\right\rvert_{\varepsilon=0}.
\end{align}

Let $\rho$ be a faithful bosonic Gaussian thermal state given by $\rho\coloneq e^{-\hat{G}}/\Tr[e^{-\hat{G}}]$, where $\hat{G}$ is Hermitian. We define $\sigma(\phi)$ as the unitary evolution of $\rho$ of the form
\begin{equation}
\sigma(\phi)\coloneq e^{-i\hat{H}(\phi)}\rho e^{i\hat{H}(\phi)},     
\end{equation}
where
\begin{align}
    \hat{H}(\phi)&\coloneq\sum_j\phi_jH_j,
\end{align}
and each $H_j$ is Hermitian. For $\alpha\in(0,1)$ and $z>0$, the entries of the $\alpha$-$z$ information matrix are given by~\cite[Theorem 35]{wilde2025}
\begin{align}
    [I_{\alpha,z}(\phi)]_{i,j}&=\ba{\,\bb{\Phi_{p_{\alpha,z}}(\Psi_\phi(H_i)),\bb{\hat{G},\Psi_\phi(H_j)}}}_{\rho}.
    \label{res:alpha_z_info_mtx_time_evo_states0}
\end{align}
Here, $\Psi_\phi$ and $\Phi_{p_{\alpha,z}}$ are quantum channels defined as
\begin{align}
    \Psi_\phi(X)&\coloneq\int_{0}^1dt\,e^{i\hat{H}(\phi)t}Xe^{-i\hat{H}(\phi)t},
    \label{def:psi_channel_time_evo}\\
    \Phi_{p_{\alpha,z}}(X)&\coloneq\int_{-\infty}^\infty dt\,p_{\alpha,z}(t)e^{-i\hat{G}t}Xe^{i\hat{G}t},
    \label{def:phi_channel_time_evo}
\end{align}
and $p_{\alpha,z}(t)$ is a probability density given by
\begin{align}
    p_{\alpha,z}(t)\coloneq\frac{z}{2\pi\alpha(1-\alpha)}\ln\bp{1+\bp{\frac{\sin(\pi\alpha)}{\sinh(\pi zt)}}^2}.
\end{align}

The presence of $H_i$ and $H_j$ in~\eqref{res:alpha_z_info_mtx_time_evo_states0} is due to the fact that $H_i=\frac{\pd}{\pd \phi_i}\hat{H}(\phi)$ when $\hat{H}(\phi)=\sum_j\phi_jH_j$. When we do not assume that $\hat{H}(\phi)$ is linear in $\phi$, the formula in~\eqref{res:alpha_z_info_mtx_time_evo_states0} can be extended to the following general formula:
\begin{multline}
    [I_{\alpha,z}(\phi)]_{i,j}\\
    =\ba{\,\bb{\Phi_{p_{\alpha,z}}(\Psi_\phi(\pd_i\hat{H}(\phi))),\bb{\hat{G},\Psi_\phi(\pd_j\hat{H}(\phi))}}}_{\rho},
    \label{res:alpha_z_info_mtx_time_evo_states}
\end{multline}
where we adopted the shorthand $\pd_i(\cdot)\coloneq\frac{\pd}{\pd\phi_i}(\cdot)$.
In particular, for our purposes here, we are interested in a quadratic Hamiltonian of the form
\begin{align}
    \hat{H}(\phi)&\coloneq\frac{1}{2}(\hat{r}^{c_r})^TH(\hat{r}^{c_r}),
\end{align}
where $H$ is a real and symmetric Hamiltonian matrix and $\hat{r}^{c_r}$ is defined as in~\eqref{def:centered_vec}. When $\alpha=z=\frac{1}{2}$, the $\alpha$-$z$ information matrix reduces to the symmetric logarithmic derivative (SLD) QFI matrix. This occurs because the $\alpha$-$z$ R\'enyi relative entropy reduces to the negative logarithm of the fidelity for $\alpha=z=\frac{1}{2}$, and it is known that this divergence leads to SLD Fisher information.

\subsection{Quantum Cr\'amer--Rao bound and SLD operator}\label{subsec:prelim_qcr}

For a family $(\sigma(\phi))_{\phi}$ of parametrized states, the multiparameter quantum Cr\'amer--Rao bound states that~\cite{helstrom1968}
\begin{align}
    \mathrm{Cov}^{(n)}(\hat{\phi},\phi)\geq\frac{1}{n}\bb{I_{\alpha,z}(\phi)}^{-1},
\end{align}
for $\alpha = z = \frac{1}{2}$ (i.e., the SLD QFI),
where $\hat{\phi}$ is an unbiased estimate of the parameter vector $\phi$, $n\in\NN$ is the number of copies of the state, and the entries of the covariance matrix $\mathrm{Cov}^{(n)}(\hat{\phi},\phi)$ are given by
\begin{multline}
    \bb{\mathrm{Cov}^{(n)}(\hat{\phi},\phi)}_{i,j}\\
    \coloneq\sum_m\Tr\bb{M_m^{(n)}\sigma(\phi)^{\otimes n}}\bp{\hat{\phi}_i(m)-\phi_i}\bp{\hat{\phi}_j(m)-\phi_j}.
\end{multline}
Here, $(M^{(n)}_m)_m$ is a positive operator-valued measure (POVM) that describes the joint measurement used for the estimation, and $m$ refers to a measurement outcome.

The SLD operator gives rise to an optimal POVM (i.e., that saturates the quantum Cr\'amer--Rao bound) for single parameter estimation via an SLD spectral measurement~\cite{braunstein_caves1994}. Let $\sigma(\phi)=\sum_{k}\lambda_k\op{k}{k}$ be the spectral decomposition of $\sigma(\phi)$. Then the SLD operator corresponding to an entry $\phi_j$ of the parameter vector $\phi$ is defined as~\cite[Eq.~(86)]{sidhu2020}
\begin{align}
    L^{(j)}(\phi)\coloneq \sum_{k,\ell}\frac{2}{\lambda_k+\lambda_\ell}\ip{k}{\pd_j\sigma(\phi)\middle|\ell}\op{k}{\ell},\label{def:sld}
\end{align}
and it satisfies the following differential equation~\cite[Eq.~(83)]{sidhu2020}:
\begin{align}
    \pd_j\sigma(\phi)=\frac{1}{2}\bc{L^{(j)}(\phi),\sigma(\phi)}.
\end{align}

\section{Results}\label{sec:res}

\subsection{\boldmath$\alpha$-$z$ information matrix for time-evolved bosonic Gaussian thermal states}\label{subsec:res_alpha_z_mtx}

In this section, we establish formulas for the entries of the $\alpha$--$z$ information matrix when the underlying family of parametrized states consists of time-evolved bosonic Gaussian thermal states.

Let $\rho$ be a faithful $n$-mode bosonic Gaussian thermal state  of the form
\begin{align}
    \rho=\frac{e^{-\hat{G}}}{Z(\mu,G)},\label{eq:rho_therm}
\end{align}
with $2n\times 1$ mean vector $\mu$ and $2n\times 2n$ Hamiltonian matrix $G$,
where
\begin{align}
    \hat{G}\coloneq\frac{1}{2}(\hat{r}^{c_\mu})^TG\hat{r}^{c_\mu},\label{def:ghat}
\end{align}
and $Z=\Tr[e^{-\hat{G}}]$. We define $\sigma(\phi)$ as the unitary time evolution of $\rho$ of the form
\begin{align}
    \sigma(\phi)&\coloneq e^{-i\hat{H}(\phi)}\rho e^{i\hat{H}(\phi)}\label{eq:sigma_state},
\end{align}
where
\begin{align}
    \hat{H}(\phi)&\coloneq\frac{1}{2}(\hat{r}^{c_r})^TH \hat{r}^{c_r}.\label{def:H_phi}
\end{align}
Here, $H$ is a real and symmetric Hamiltonian matrix, $\hat{r}^{c_r}$ is defined as in~\eqref{def:centered_vec}, and $r\in\RR^{2n}$. We note that the time dependence of the evolution has been implicitly absorbed into the Hamiltonian matrix $H$. Additionally, an evolution of this form involves only Gaussian operations and thus preserves the Gaussianity of $\rho$. The parameter vector $\phi$ consists of the entries of the displacement vector $r$ and the Hamiltonian matrix $H$. Following the approach of~\cite{huang_wilde2024}, we define $\phi$ as
\begin{multline}
    \phi\coloneq(r_1,\ldots,r_{2n},h_{1,1},h_{1,2},\ldots,\\
    h_{1,2n},h_{2,2},h_{2,3},\ldots,h_{2n,2n}),
\end{multline}
where $h_{j,k}\coloneq[H]_{j,k}$. Because $H$ is symmetric, we have excluded redundant entries of the Hamiltonian matrix in our definition of the parameter vector (e.g. we include $h_{1,2}$ but not $h_{2,1}$). \\

Due to the structure of $\hat{H}(\phi)$ in~\eqref{def:H_phi}, the formulas for the derivatives of $\hat{H}(\phi)$ are the same as those in~\cite[Eqs.~(55) and (56)]{huang_wilde2024}. Namely,
\begin{align}
    \pd_i\hat{H}(\phi)\coloneq\frac{\pd}{\pd r_i}\hat{H}(\phi)&=-\sum_{j=1}^{2n}h_{i,j}\hat{r}_j^{c_r},\label{eq:pd_H_ri}\\
    \pd_{(j,k)}\hat{H}(\phi)\coloneq\frac{\pd}{\pd h_{j,k}}\hat{H}(\phi)&=\frac{1}{4}\bc{\hat{r}_j^{c_r},\hat{r}_k^{c_r}}\label{eq:pd_H_hjk}.
\end{align}

We now state our main result, where we establish formulas for the entries of the $\alpha$-$z$ information matrix for time-evolved faithful bosonic Gaussian thermal states of the form in~\eqref{eq:sigma_state}. In preparation for this result, we define the following matrices and vector:
\begin{align}
    A(t,s_1,s_2)&\coloneq HS(s_1)T(t)\Omega G\bp{HS(s_2)\Omega}^T,\\
    B_1(t,s_1,s_2)&\coloneq S(s_1)T(t)\Omega S^T(s_2), \label{def:B1}\\
    B_2(t,s_1,s_2)&\coloneq S(s_1)T(t)VG\Omega S^T(s_2), \label{def:B2}\\
    B_3(t,s_1,s_2)&\coloneq S(s_1)T(t)\Omega G\Omega S^T(s_2), \label{def:B3}\\
    B_4(t,s_1,s_2)&\coloneq S(s_1)T(t)VS^T(s_2), \label{def:B4}\\
    b(s)&\coloneq S(s)(\mu-r), \label{def:bvec}\\
    C(t,s_1,s_2)&\coloneq HS(s_1)T(t)\Omega G\Omega S^T(s_2),
\end{align}
where $T(t)\coloneq e^{-\Omega Gt}$ and $S(s)\coloneq e^{\Omega Hs}$.

\begin{theorem}
\label{thm:main}
Let $\rho$ be a faithful bosonic Gaussian thermal state with mean vector $\mu$ and Hamiltonian matrix $G$ as in~\eqref{eq:rho_therm}, and let $\sigma(\phi)$ be a unitary Gaussian time-evolution of $\rho$ as in~\eqref{eq:sigma_state} parametrized by a displacement vector $r$ and Hamiltonian matrix $H$. Then, for $\alpha\in(0,1)$ and $z>0$, the entries of the $\alpha$-$z$ information matrix are
\begin{widetext}
\begin{align}
    [I_{\alpha,z}(r,H)]_{i,j}&=\int_{-\infty}^\infty dt\,p_{\alpha,z}(t)\lambda_{i,j}(t),\label{eq:final_info_mtx_1}\\
    [I_{\alpha,z}(r,H)]_{(j_1,k_1),(j_2,k_2)}&=\int_{-\infty}^\infty dt\,p_{\alpha,z}(t)\mu_{j_1,k_1,j_2,k_2}(t), \label{eq:final_info_mtx_2}\\
     [I_{\alpha, z}(r, H)]_{i,(j,k)}=[I_{\alpha, z}(r, H)]_{(j,k),i}&=\int_{-\infty}^\infty dt\,p_{\alpha,z}(t)\nu_{i,j,k}(t), \label{eq:final_info_mtx_3}
\end{align}
where
\begin{align}
    \lambda_{i,j}(t)&\coloneq\int_0^1\int_0^1ds_1\, ds_2\,\bb{A}_{i,j},\\
    \mu_{j_1,k_1,j_2,k_2}(t)&\coloneq-\frac{1}{4}\int_0^1\int_0^1ds_1\, ds_2\,\sum_{\substack{(a,b)=(j_1,k_1),(k_1,j_1),\\(c,d)=(j_2,k_2),(k_2,j_2)}}\bb{B_1}_{a,c}\bb{B_2}_{b,d}+\bb{B_3}_{a,c}\bb{B_4}_{b,d} +b(s_1)_{a}\bb{B_3}_{b,d}b(s_2)_{c},\label{eq:g_alphaz_two_hamiltonian}\\
    \nu_{i,j,k}(t)&\coloneq \frac{1}{2}\int_0^1\int_0^1ds_1\, ds_2\,\bb{C}_{i,j}b(s_2)_k+\bb{C}_{i,k}b(s_2)_j,
\end{align}
\end{widetext}
and we have left the dependence of $A$, $B_1$, $B_2$, $B_3$, $B_4$, and $C$ on $s_1$, $s_2$, and $t$ implicit.

\end{theorem}

\begin{proof}
See Appendix~\ref{app:thm-proof}.
\end{proof}

\subsection{SLD for time-evolved bosonic Gaussian thermal states}\label{subsec:res_sld}
Here, we establish a formula for the SLD operator for time-evolved bosonic Gaussian thermal states. Let $\sigma(\phi)$ be a time-evolved bosonic Gaussian thermal state as defined in~\eqref{eq:sigma_state}, and let $\Psi_\phi$ denote the quantum channel defined in~\eqref{def:psi_channel_time_evo}. For a Hermitian operator $A$, we denote by $\Phi_{p,A}$ the quantum channel
\begin{align}
    \Phi_{p,A}(X)&\coloneq\int_{\RR}dt\,p(t)e^{-iAt}Xe^{iAt},\\
    p(t)&\coloneq\frac{2}{\pi}\ln\abs{\coth\bp{\frac{\pi t}{2}}}\label{def:high_peak_tent_pd},
\end{align}
where $p(t)$ is the so-called high-peak tent probability density~\cite{patel2025}.

\begin{proposition}\label{prop:sld}
    The SLD operator $L^{(j)}(\phi)$ for the parametrized family $(\sigma(\phi))_\phi$ is
    \begin{align}
        L^{(j)}(\phi)=i\bb{\Phi_{p,\hat{G}_\phi}\bp{\Psi^\dag_{\phi}\bp{\pd_j(\hat{H}(\phi))}},\hat{G}_\phi},
    \end{align}
    where $\hat{G}_\phi\coloneq e^{-i\hat{H}(\phi)}\hat{G}e^{i\hat{H}(\phi)}$ and $\Psi^\dag_\phi$ is the adjoint of the channel in~\eqref{def:psi_channel_time_evo}.
\end{proposition}
\begin{proof}
    See Appendix~\ref{app:sld}.
\end{proof}

\subsection{Parameter estimation for a beam splitter}\label{subsec:res_ex}
A beam splitting operation is characterized by parameters $\theta\in[0,\pi/2]$ and $\ph\in[0,2\pi]$, where $\eta\coloneq(\cos\theta)^2$ is the transmissivity of the beam splitter, satisfying $0\leq\eta\leq1$, and $\ph$ corresponds to a local phase shift~\cite{serafini2017}. The Hamiltonian matrix for a beam splitting transformation is given by~\cite{serafini2017}
\begin{align}
    H_{BS}(\theta,\ph)&=\theta\begin{bmatrix}
        \mathbf{0}&D(\ph)\\D(\ph)^T&\mathbf{0}
    \end{bmatrix},\\
    D(\ph)&\coloneq\begin{bmatrix}
        \cos\ph&-\sin\ph\\
        \sin\ph&\cos\ph
    \end{bmatrix}.\label{def:d_ph}
\end{align}
The corresponding symplectic transformation is
\begin{align}
    e^{\Omega H_{BS}}&=\begin{bmatrix}
        \cos\theta\ii_2 & \sin\theta R(\ph)\\
        -\sin\theta R^T(\ph) & \cos\theta\ii_2
    \end{bmatrix},\\
    R(\ph)&\coloneq\begin{bmatrix}
        \sin\ph & \cos\ph \\
        -\cos\ph & \sin\ph
    \end{bmatrix}.
\end{align}
We note that the operator $\hat{H}$ for this transformation is
\begin{align}
    \hat{H}=\frac{1}{2}(\hat{r})^TH_{BS}\hat{r}.
\end{align}
The displacement vector here is equal to zero, and thus the parameter vector $\phi$ consists solely of the entries of the Hamiltonian matrix $H_{BS}$.

\begin{figure}
    \centering
    \includegraphics[width=1.0\linewidth]{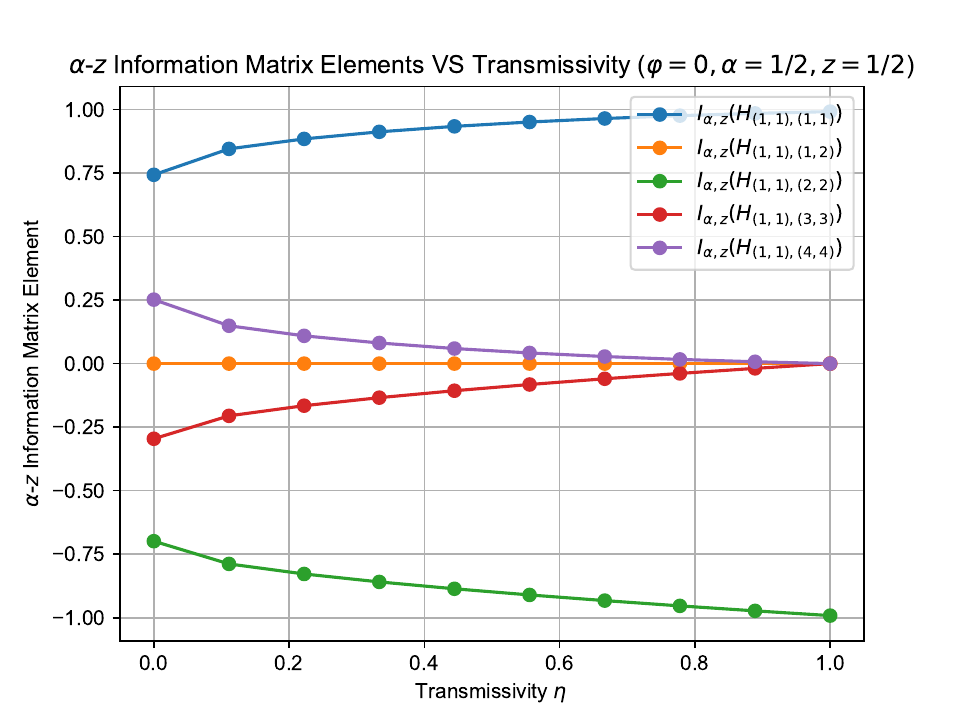}
    \caption{Select elements of the $\alpha$-$z$ information matrix are plotted versus the transmissivity $\eta$, where $\ph=0$, $\alpha=1/2$, $z=1/2$, $\overline{n}_1=5$, and $\overline{n}_2=10$.}
    \label{fig:phi0_infomtx}
\end{figure}

We consider a two-mode thermal state as the input to the beam splitter, where the modes have mean photon numbers $\overline{n}_1>0$ and $\overline{n}_2>0$, respectively. In this scenario, the Hamiltonian and covariance matrices are~\cite{serafini2017} 
\begin{align}
    G&=\lambda_1\ii_2\oplus\lambda_2\ii_2\\
    &=\begin{bmatrix}
        \lambda_1&0\\0&\lambda_2
    \end{bmatrix}\otimes\ii_2,\\
    V&=\coth\bp{\frac{\lambda_1}{2}}\frac{\ii_2}{2}\oplus\coth\bp{\frac{\lambda_2}{2}}\frac{\ii_2}{2}\\
    &=\begin{bmatrix}
        \coth(\lambda_1/2)&0\\0&\coth(\lambda_2/2)
    \end{bmatrix}\otimes\frac{\ii_2}{2},
\end{align}
where $\coth(\lambda_i/2)=2\overline{n}_i+1$ for $i=1,2$. In Fig.~\ref{fig:phi0_infomtx}, we plot various entries of the $\alpha$-$z$ information matrix versus the transmissivity $\eta$ when $\overline{n}_1=5$, $\overline{n}_2=10$, $\ph=0$, $\alpha=1/2$, and $z=1/2$. 

We note that the plotted information matrix entries in Fig.~\ref{fig:phi0_infomtx} correspond to entries of the Hamiltonian matrix, and information matrix entries corresponding to the parameters $\theta$ and $\ph$ may be obtained by an appropriate Jacobian transformation. The Jacobian transformation is given by~\cite[Eq.~(28)]{sidhu2020}
\begin{multline}
    \bb{I_{\alpha,z}(\theta,\ph)}_{i,j}=\sum_{\substack{k_1,k_2,\\\ell_1,\ell_2}=1}^4\pd_iH_{BS,k_1,\ell_1}(\theta,\ph)\times\\
    I_{\alpha,z}(H_{BS}(\theta,\ph))_{(k_1,\ell_1),(k_2,\ell_2)}\pd_j H_{BS,k_2,\ell_2}(\theta,\ph).
\end{multline}
Let us find the derivatives of $H_{BS}$ with respect to $\theta$ and $\ph$. The derivative with respect to $\theta$ is
\begin{align}
    \pd_\theta H_{BS}(\theta,\ph)&=\pd_\theta \bp{\theta F(\ph)}\\
    &=F(\ph),
\end{align}
where
\begin{align}
    F(\ph)\coloneq\begin{bmatrix}
        \bm{0}&D(\ph)\\ D(\ph)^T&\bm{0}
    \end{bmatrix},
\end{align}
and $D(\ph)$ is defined in~\eqref{def:d_ph}.
The derivative with respect to $\ph$ is
\begin{align}
    \pd_\ph H_{BS}(\theta,\ph)&=\theta\begin{bmatrix}
        \bm{0}&\pd_\ph D(\ph)\\ \pd_\ph D(\ph)^T&\bm{0}
    \end{bmatrix}.
\end{align}
Consider that
\begin{align}
    \pd_\ph D(\ph)&=\begin{bmatrix}
        -\sin\ph&-\cos\ph\\ \cos\ph&-\sin\ph
    \end{bmatrix}\\
    &=D(\ph+\pi/2).
\end{align}
Similarly, $\pd_\ph D(\ph)^T=D(\ph+\pi/2)^T$.
Then, we can write $\pd_\ph H_{BS}(\theta,\ph)$ as
\begin{align}
    \pd_\ph H_{BS}(\theta,\ph)&=H_{BS}(\theta,\ph+\pi/2).
\end{align}
It follows that
\begin{widetext}
\begin{align}
    \bb{I_{\alpha,z}(\theta,\ph)}_{\theta,\theta}&=\sum_{\substack{k_1,k_2,\\\ell_1,\ell_2}=1}^4F_{k_1,\ell_1}(\ph)I_{\alpha,z}(H_{BS}(\theta,\ph))_{(k_1,\ell_1),(k_2,\ell_2)}F_{k_2,\ell_2}(\ph),\\
    \bb{I_{\alpha,z}(\theta,\ph)}_{\ph,\ph}&=\sum_{\substack{k_1,k_2,\\\ell_1,\ell_2}=1}^4H_{BS,k_1,\ell_1}(\theta,\ph+\pi/2)I_{\alpha,z}(H_{BS}(\theta,\ph))_{(k_1,\ell_1),(k_2,\ell_2)}H_{BS,k_2,\ell_2}(\theta,\ph+\pi/2),\\
    \bb{I_{\alpha,z}(\theta,\ph)}_{\theta,\ph}=\bb{I_{\alpha,z}(\theta,\ph)}_{\ph,\theta}&=\sum_{\substack{k_1,k_2,\\\ell_1,\ell_2}=1}^4F_{k_1,\ell_1}(\ph)I_{\alpha,z}(H_{BS}(\theta,\ph))_{(k_1,\ell_1),(k_2,\ell_2)}H_{BS,k_2,\ell_2}(\theta,\ph+\pi/2).
\end{align}
\end{widetext}

\section{Conclusion}\label{sec:concl}
In this work, we considered the task of estimating parameters that govern a Gaussian unitary evolution of a bosonic Gaussian thermal state. For a parameterized family of these time-evolved bosonic Gaussian thermal states, we established formulas for the entries of the $\alpha$-$z$ information matrix, which is the QFI matrix arising from the $\alpha$-$z$ R\'enyi relative entropy. Furthermore, we found a formula for the SLD operator for time-evolved bosonic Gaussian states. As an application, we calculated entries of the $\alpha$-$z$ information matrix when the initial state is a two-mode thermal state, and the evolution is a beam splitter.

Our results have many possible applications in quantum metrology. As an extension to the example of parameter estimation for a single beam splitter, one could consider parameter estimation for an interferometer consisting of an array of beam splitters. 
Furthermore, our results could be used for the estimation of parameters describing displacements or squeezing transformations as well as coupling rates between modes of bosonic Gaussian states. Another future direction could be finding a formula for the QFI of time-evolved fermionic Gaussian states.

\textit{Note on independent work: After completing the technical development of this work, the following related recent work appeared on the arXiv~\cite{chitgopekar2026}. The main contributions of our paper, Theorem~\ref{thm:main} and Proposition~\ref{prop:sld}, are distinct from those of~\cite{chitgopekar2026}.}

\section{Acknowledgements}
We thank Zixin Huang for helpful discussions. HSM~acknowledges support from NSF grant No.~DGE-2139899. This work was also supported
by the Quantum Science Center (QSC), a National Quantum Information Science Research Center of the U.S.~Department of Energy (DOE).

\bibliographystyle{unsrturl}
\bibliography{references}
\onecolumngrid
\appendix

\section{Proof of Theorem~\ref{thm:main}}

\label{app:thm-proof}

Before providing a proof of Theorem~\ref{thm:main}, we state the following lemma, which involves the action of a quadratic Hamiltonian on quadrature operators.
\begin{lemma}
    Let $\hat{G}$ be defined as in~\eqref{def:ghat}, and let $\hat{r}^{c_\mu}$ and $\hat{r}^{c_r}$ be defined as in~\eqref{def:centered_vec}. Then,
    \begin{align}
        e^{i\hat{G}t}\hat{r}^{c_r}_ke^{-i\hat{G}t}
        &=\bb{T(t)\hat{r}^{c_\mu}-r+\mu}_k,
    \end{align}
    where $T(t)\coloneq e^{\Omega Gt}$.
    \label{lem:hadamard_diff_mean_vecs}
\end{lemma}
\begin{proof}
    We first write $\hat{r}^{c_r}$ as $\hat{r}^{c_\mu}-r+\mu$. Then,
    \begin{align}
        e^{i\hat{G}t}\hat{r}^{c_r}_ke^{-i\hat{G}t}&=e^{i\hat{G}t}\hat{r}^{c_\mu}_ke^{-i\hat{G}t}-r_k+\mu_k\\
        &=\bp{\sum_{\ell=1}^{2n}T_{k,\ell}(t)\hat{r}^{c_\mu}_\ell}-r_k+\mu_k \label{lem:hadamard1}\\
        &=\bb{T(t)\hat{r}^{c_\mu}-r+\mu}_k,
    \end{align}
    where we used~\cite[Lemma 1]{huang_wilde2024} to arrive at~\eqref{lem:hadamard1}.
\end{proof}

We now provide a proof of Theorem~\ref{thm:main}.
We will proceed as follows: First, we will find expressions for the action of the channel $\Psi_\phi$ in \eqref{def:psi_channel_time_evo} on arbitrary $\pd_i\hat{H}(\phi)$ and $\pd_{(j,k)}\hat{H}(\phi)$, and we will then find expressions for the action of $\Phi_{p_{\alpha,z}}$ on the results. Then, we will find expressions for the commutators of $\hat{G}$ with $\Psi_\phi(\pd_i\hat{H}(\phi))$ and $\Psi_\phi(\pd_{(j,k)}\hat{H}(\phi))$. After this, we will be in a position to derive the final formulas in~\eqref{eq:final_info_mtx_1}--\eqref{eq:final_info_mtx_3}.

Let us begin by finding $\Psi_{r,H}(\pd_i\hat{H}(\phi))$ and $\Psi_{r,H}(\pd_{(j,k)}\hat{H}(\phi))$. The action of the channel $\Psi_{r,H}$, given in \eqref{def:psi_channel_time_evo}, on $\pd_i\hat{H}(\phi)$ is
\begin{align}
    \Psi_{r,H}(\pd_i\hat{H}(\phi))&=\int_0^1ds\,e^{i\hat{H}(\phi)s}\pd_i\hat{H}(\phi)e^{-i\hat{H}(\phi)s}\\
    &=-\int_0^1ds\,\sum_{m_1=1}^{2n}h_{i,m_1}e^{i\hat{H}(\phi)s}\hat{r}_{m_1}^{c_r}e^{-i\hat{H}(\phi)s},
    \label{intres:psi_channel_action_dis_vec}
\end{align}
where we used~\eqref{eq:pd_H_ri} to arrive at~\eqref{intres:psi_channel_action_dis_vec}. Defining $S(s)\coloneq e^{\Omega Hs}$, we conclude from Lemma~\ref{lem:hadamard_diff_mean_vecs} that
\begin{align}
    e^{i\hat{H}(\phi)s}\hat{r}_i^{c_r}e^{-i\hat{H}(\phi)s}&=\sum_{m=1}^{2n}S_{i,m}(s)\hat{r}^{c_r}_m.
\end{align}
With this, Eq.~\eqref{intres:psi_channel_action_dis_vec} becomes
\begin{align}
    \Psi_{r,H}(\pd_i\hat{H}(\phi))&=-\sum_{m_1=1}^{2n}\sum_{m_2=1}^{2n}\int_0^1ds\,h_{i,m_1}S_{m_1,m_2}(s)\hat{r}^{c_r}_{m_2}.
    \label{res:psi_H_i}
\end{align}
Now, we will find $\Psi_{r,H}(\pd_{(j,k)}\hat{H}(\phi))$: 
\begin{align}
     \Psi_{r,H}(\pd_{(j,k)}\hat{H}(\phi))&=\int_0^1ds\,e^{i\hat{H}(\phi)s}\pd_{(j,k)}\hat{H}(\phi)e^{-i\hat{H}(\phi)s}\\
     &=\frac{1}{4}\int_0^1ds\,e^{i\hat{H}(\phi)s}\bc{\hat{r}_j^{c_r},\hat{r}_k^{c_r}}e^{-i\hat{H}(\phi)s} \label{eq:psi_pd_jk_H1}\\
     &=\frac{1}{4}\int_0^1ds\,\bc{e^{i\hat{H}(\phi)s}\hat{r}_j^{c_r}e^{-i\hat{H}(\phi)s},e^{i\hat{H}(\phi)s}\hat{r}_k^{c_r}e^{-i\hat{H}(\phi)s}}\\
     &=\frac{1}{4}\sum_{m_1=1}^{2n}\sum_{m_2=1}^{2n}\int_0^1ds\,S_{j,m_1}(s)S_{k,m_2}(s)\bc{\hat{r}^{c_r}_{m_1},\hat{r}^{c_r}_{m_2}},
     \label{res:psi_H_jk}
\end{align}
where we used~\eqref{eq:pd_H_hjk} in~\eqref{eq:psi_pd_jk_H1} and Lemma~\ref{lem:hadamard_diff_mean_vecs} in~\eqref{res:psi_H_jk}. Now, let us find expressions for $\Phi_{p_{\alpha,z}}(\Psi_{r,H}(\pd_i\hat{H}(\phi)))$ and $\Phi_{p_{\alpha,z}}(\Psi_{r,H}(\pd_{(j,k)}\hat{H}(\phi)))$. Consider that
\begin{align}
    & \Phi_{p_{\alpha,z}}(\Psi_{r,H}(\pd_i\hat{H}(\phi)))\notag \\
    &=\int_{-\infty}^\infty dt\,p_{\alpha,z}(t)e^{-i\hat{G}t}\Psi_{r,H}(\pd_i\hat{H}(\phi))e^{i\hat{G}t}\\
    &=-\sum_{m_1=1}^{2n}\sum_{m_2=1}^{2n}\int_0^1ds\,h_{i,m_1}S_{m_1,m_2}(s)\int_{-\infty}^\infty dt\,p_{\alpha,z}(t)e^{-i\hat{G}t}\hat{r}^{c_r}_{m_2}e^{i\hat{G}t}\\
    &=-\sum_{m_1=1}^{2n}\sum_{m_2=1}^{2n}\int_0^1ds\,h_{i,m_1}S_{m_1,m_2}(s)\int_{-\infty}^\infty dt\,p_{\alpha,z}(t)\left(\bp{\sum_{\ell=1}^{2n}T_{m_2,\ell}(t)\hat{r}^{c_\mu}_\ell}+\mu_{m_2}-r_{m_2}\right), \label{eq:phi_psi_H_i}
\end{align}
where we used \eqref{res:psi_H_i},
and we also employed Lemma \ref{lem:hadamard_diff_mean_vecs} with $t\ra-t$. We note that, from here on out, we define $T(t)\coloneq e^{-\Omega Gt}$. We will now consider $\Phi_{p_{\alpha,z}}(\Psi_{r,H}(\pd_{(j,k)}\hat{H}(\phi)))$:
\begin{align}
    \Phi_{p_{\alpha,z}}&(\Psi_{r,H}(\pd_{(j,k)}\hat{H}(\phi))) \notag\\
    &=\int_{-\infty}^\infty dt\,p_{\alpha,z}(t)e^{-i\hat{G}t}\Psi_{r,H}(\pd_{(j,k)}\hat{H}(\phi))e^{i\hat{G}t}\\
    &=\frac{1}{4}\sum_{m_1=1}^{2n}\sum_{m_2=1}^{2n}\int_0^1ds\,S_{j,m_1}(s)S_{k,m_2}(s)\int_{-\infty}^\infty dt\,p_{\alpha,z}(t)e^{-i\hat{G}t}\bc{\hat{r}^{c_r}_{m_1},\hat{r}^{c_r}_{m_2}}e^{i\hat{G}t}\\
    &=\frac{1}{4}\sum_{m_1=1}^{2n}\sum_{m_2=1}^{2n}\int_0^1ds\,S_{j,m_1}(s)S_{k,m_2}(s)\int_{-\infty}^\infty dt\,p_{\alpha,z}(t)\bc{e^{-i\hat{G}t}\hat{r}^{c_r}_{m_1}e^{i\hat{G}t},e^{-i\hat{G}t}\hat{r}^{c_r}_{m_2}e^{i\hat{G}t}}\\
    &=\frac{1}{4}\sum_{m_1,m_2=1}^{2n}\int_0^1ds\,S_{j,m_1}(s)S_{k,m_2}(s)\int_{-\infty}^\infty dt\,p_{\alpha,z}(t)\left\{\bp{\sum_{\ell_1=1}^{2n}T_{m_1,\ell_1}(t)\hat{r}^{c_\mu}_{\ell_1}}+(\mu-r)_{m_1},\right. \notag\\
    &\hspace{8.4cm}\left.\bp{\sum_{\ell_2=1}^{2n}T_{m_2,\ell_2}(t)\hat{r}^{c_\mu}_{\ell_2}}+(\mu-r)_{m_2}\right\}\\
    &=\frac{1}{4}\sum_{m_1,m_2=1}^{2n}\int_{-\infty}^\infty dt\int_0^1ds\,p_{\alpha,z}(t)S_{j,m_1}(s)S_{k,m_2}(s)\left(\sum_{\ell_1,\ell_2=1}^{2n}T_{m_1,\ell_1}(t)T_{m_2,\ell_2}(t)\bc{\hat{r}^{c_\mu}_{\ell_1},\hat{r}^{c_\mu}_{\ell_2}}\right. \notag\\
    &\hspace{.5cm}\left.+2\sum_{\ell_1=1}^{2n}T_{m_1,\ell_1}(t)\hat{r}^{c_\mu}_{\ell_1}(\mu-r)_{m_2}+2\sum_{\ell_2=1}^{2n}T_{m_2,\ell_2}(t)\hat{r}^{c_\mu}_{\ell_2}(\mu-r)_{m_1}+2(\mu-r)_{m_1}(\mu-r)_{m_2}\right), \label{eq:phi_psi_H_jk}
\end{align}
where we used \eqref{res:psi_H_jk} and Lemma \ref{lem:hadamard_diff_mean_vecs}. Our next step will be to find the commutators $\bb{\hat{G},\Psi_{r,H}(\pd_i\hat{H}(\phi))}$ and $\bb{\hat{G},\Psi_{r,H}(\pd_{(j,k)}\hat{H}(\phi))}$. First, note that
\begin{align}
        \bb{\hat{r}^{c_\mu}_j,\hat{r}^{c_r}_k}&=\bb{\hat{r}_j-\mu_j,\hat{r}_k-r_k}\\
        &=\bb{\hat{r}_j,\hat{r_k}}-\bb{\hat{r}_j,r_k}-\bb{\mu_j,\hat{r}_k}+\bb{\mu_j,r_k}\\
        &=\bb{\hat{r}_j,\hat{r_k}}\\
        &=i\Omega_{j,k}.
        \label{res:commutator_diff_mean_vecs}
    \end{align}
    It follows that
\begin{align}
        \bb{\hat{r}^{c_\mu}_i\hat{r}^{c_\mu}_j,\hat{r}^{c_r}_k}&=\hat{r}^{c_\mu}_i\hat{r}^{c_\mu}_j\hat{r}^{c_r}_k-\hat{r}^{c_r}_k\hat{r}^{c_\mu}_i\hat{r}^{c_\mu}_j\\
        &=\hat{r}^{c_\mu}_i\hat{r}^{c_\mu}_j\hat{r}^{c_r}_k-\hat{r}^{c_\mu}_i\hat{r}^{c_r}_k\hat{r}^{c_\mu}_j+\hat{r}^{c_\mu}_i\hat{r}^{c_r}_k\hat{r}^{c_\mu}_j-\hat{r}^{c_r}_k\hat{r}^{c_\mu}_i\hat{r}^{c_\mu}_j\\
        &=\hat{r}^{c_\mu}_i\bb{\hat{r}^{c_\mu}_j,\hat{r}^{c_r}_k}+\bb{\hat{r}^{c_\mu}_i,\hat{r}^{c_r}_k}\hat{r}^{c_\mu}_j\\
        &=i\bp{\hat{r}^{c_\mu}_i\Omega_{j,k}+\hat{r}^{c_\mu}_j\Omega_{i,k}}.
        \label{res:commutator_2quadops_1quadop_diff_mean_vecs}
    \end{align}
Recalling that $\hat{G}=\frac{1}{2}(\hat{r}^{c_\mu})^TG(\hat{r}^{c_\mu})=\frac{1}{2}\sum_{n_1,n_2=1}^{2n}G_{n_1,n_2}\hat{r}^{c_\mu}_{n_1}\hat{r}^{c_\mu}_{n_2}$, we can use \eqref{res:psi_H_i} and \eqref{res:commutator_2quadops_1quadop_diff_mean_vecs} to find that
\begin{align}
    \bb{\hat{G},\Psi_{r,H}(\pd_i\hat{H}(\phi))}&=-\frac{1}{2}\sum_{n_1,n_2=1}^{2n}\sum_{m_1,m_2=1}^{2n}G_{n_1,n_2}\int_0^1ds\,h_{i,m_1}S_{m_1,m_2}(s)\bb{\hat{r}^{c_\mu}_{n_1}\hat{r}^{c_\mu}_{n_2},\hat{r}^{c_r}_{m_2}}\\
    &=-\frac{1}{2}\sum_{n_1,n_2,m_1,m_2=1}^{2n}G_{n_1,n_2}\int_0^1ds\,h_{i,m_1}S_{m_1,m_2}(s)\bp{\hat{r}^{c_\mu}_{n_1}i\Omega_{n_2,m_2}+\hat{r}^{c_\mu}_{n_2}i\Omega_{n_1,m_2}}. \label{eq:commutator_g_psi_h_i}
\end{align}
We will now proceed to find $\bb{\hat{G},\Psi_{r,H}(\pd_{(j,k)}\hat{H}(\phi))}$. Using \eqref{res:psi_H_jk}, it follows that
\begin{align}
    \bb{\hat{G},\Psi_{r,H}(\pd_{(j,k)}\hat{H}(\phi))}&=\frac{1}{8}\sum_{m_1,m_2=1}^{2n}\sum_{n_1,n_2=1}^{2n}G_{n_1,n_2}\int_0^1ds\,S_{j,m_1}(s)S_{k,m_2}(s)\bb{\hat{r}^{c_\mu}_{n_1}\hat{r}^{c_\mu}_{n_2},\bc{\hat{r}^{c_r}_{m_1},\hat{r}^{c_r}_{m_2}}}.
    \label{intres:commutator_g_psi_h_jk}
\end{align}
In order to simplify $\bb{\hat{r}^{c_\mu}_{n_1}\hat{r}^{c_\mu}_{n_2},\bc{\hat{r}^{c_r}_{m_1},\hat{r}^{c_r}_{m_2}}}$, let us first consider a more general expression. Using \eqref{res:commutator_diff_mean_vecs}, we find that $\hat{r}^{c_a}_i\hat{r}^{c_b}_j=i\Omega_{i,j}+\hat{r}^{c_b}_j\hat{r}^{c_a}_i$. Consider the following:
\begin{align}
    \bb{\hat{r}^{c_a}_i\hat{r}^{c_b}_j,\hat{r}^{c_c}_k\hat{r}^{c_d}_l}&=\hat{r}^{c_a}_i\hat{r}^{c_b}_j\hat{r}^{c_c}_k\hat{r}^{c_d}_l-\hat{r}^{c_c}_k\hat{r}^{c_d}_l\hat{r}^{c_a}_i\hat{r}^{c_b}_j\\
    &=\hat{r}^{c_a}_i\hat{r}^{c_c}_k\hat{r}^{c_b}_j\hat{r}^{c_d}_l+i\Omega_{j,k}\hat{r}^{c_a}_i\hat{r}^{c_d}_l-\hat{r}^{c_c}_k\hat{r}^{c_d}_l\hat{r}^{c_a}_i\hat{r}^{c_b}_j\\
    &=\hat{r}^{c_c}_k\hat{r}^{c_a}_i\hat{r}^{c_b}_j\hat{r}^{c_d}_l+i\Omega_{i,k}\hat{r}^{c_b}_j\hat{r}^{c_d}_l+i\Omega_{j,k}\hat{r}^{c_a}_i\hat{r}^{c_d}_l-\hat{r}^{c_c}_k\hat{r}^{c_d}_l\hat{r}^{c_a}_i\hat{r}^{c_b}_j\\
    &=\hat{r}^{c_c}_k\hat{r}^{c_a}_i\hat{r}^{c_d}_l\hat{r}^{c_b}_j+i\Omega_{j,l}\hat{r}^{c_c}_k\hat{r}^{c_a}_i+i\Omega_{i,k}\hat{r}^{c_b}_j\hat{r}^{c_d}_l+i\Omega_{j,k}\hat{r}^{c_a}_i\hat{r}^{c_d}_l-\hat{r}^{c_c}_k\hat{r}^{c_d}_l\hat{r}^{c_a}_i\hat{r}^{c_b}_j\\
    &=i\Omega_{i,l}\hat{r}^{c_c}_k\hat{r}^{c_b}_j+i\Omega_{j,l}\hat{r}^{c_c}_k\hat{r}^{c_a}_i+i\Omega_{i,k}\hat{r}^{c_b}_j\hat{r}^{c_d}_l+i\Omega_{j,k}\hat{r}^{c_a}_i\hat{r}^{c_d}_l.
\end{align}
It follows that
\begin{align}
    \bb{\hat{r}^{c_a}_i\hat{r}^{c_b}_j,\bc{\hat{r}^{c_c}_k,\hat{r}^{c_d}_l}} & =i\Omega_{i,l}\hat{r}^{c_c}_k\hat{r}^{c_b}_j+i\Omega_{j,l}\hat{r}^{c_c}_k\hat{r}^{c_a}_i+i\Omega_{i,k}\hat{r}^{c_b}_j\hat{r}^{c_d}_l+i\Omega_{j,k}\hat{r}^{c_a}_i\hat{r}^{c_d}_l \notag\\
    &\qquad +i\Omega_{i,k}\hat{r}^{c_d}_l\hat{r}^{c_b}_j+i\Omega_{j,k}\hat{r}^{c_d}_l\hat{r}^{c_a}_i+i\Omega_{i,l}\hat{r}^{c_b}_j\hat{r}^{c_c}_k+i\Omega_{j,l}\hat{r}^{c_a}_i\hat{r}^{c_c}_k\\
    & = i\Omega_{i,l}\bc{\hat{r}^{c_c}_k,\hat{r}^{c_b}_j}+i\Omega_{j,l}\bc{\hat{r}^{c_c}_k,\hat{r}^{c_a}_i}+i\Omega_{i,k}\bc{\hat{r}^{c_b}_j,\hat{r}^{c_d}_l}+i\Omega_{j,k}\bc{\hat{r}^{c_a}_i,\hat{r}^{c_d}_l}.
    \label{res:commutator_2quad_anticomm_2quad}
\end{align}
From this, we conclude that
\begin{align}
    \bb{\hat{r}^{c_\mu}_{n_1}\hat{r}^{c_\mu}_{n_2},\bc{\hat{r}^{c_r}_{m_1},\hat{r}^{c_r}_{m_2}}}
    =i\Omega_{n_1,m_2}\bc{\hat{r}^{c_r}_{m_1},\hat{r}^{c_\mu}_{n_2}}+i\Omega_{n_2,m_2}\bc{\hat{r}^{c_r}_{m_1},\hat{r}^{c_\mu}_{n_1}}+i\Omega_{n_1,m_1}\bc{\hat{r}^{c_\mu}_{n_2},\hat{r}^{c_r}_{m_2}}+i\Omega_{n_2,m_1}\bc{\hat{r}^{c_\mu}_{n_1},\hat{r}^{c_r}_{m_2}}.
\end{align}
We now return to~\eqref{intres:commutator_g_psi_h_jk} and find that
\begin{multline}
    \bb{\hat{G},\Psi_{r,H}(\pd_{(j,k)}\hat{H}(\phi))}=\frac{1}{8}\sum_{m_1,m_2=1}^{2n}\sum_{n_1,n_2=1}^{2n}G_{n_1,n_2}\int_0^1ds\,S_{j,m_1}(s)S_{k,m_2}(s)\, \times \\ \bigg(i\Omega_{n_1,m_2}\bc{\hat{r}^{c_r}_{m_1},\hat{r}^{c_\mu}_{n_2}}
    +i\Omega_{n_2,m_2}\bc{\hat{r}^{c_r}_{m_1},\hat{r}^{c_\mu}_{n_1}}+i\Omega_{n_1,m_1}\bc{\hat{r}^{c_\mu}_{n_2},\hat{r}^{c_r}_{m_2}}+i\Omega_{n_2,m_1}\bc{\hat{r}^{c_\mu}_{n_1},\hat{r}^{c_r}_{m_2}}\bigg). \label{eq:commutator_g_psi_h_jk}
\end{multline}
We now turn our focus to finding an expression for the commutator $\bb{\Phi_{p_{\alpha,z}}(\Psi_\phi(\pd_i\hat{H}(\phi))),\bb{\hat{G},\Psi_\phi(\pd_j\hat{H}(\phi))}}$. Consider the following:
\begin{align}
    \big[\Phi_{p_{\alpha,z}}&(\Psi_\phi(\pd_i\hat{H}(\phi))),\bb{\hat{G},\Psi_\phi(\pd_j\hat{H}(\phi))}\big] \notag\\
    &=\left[-\sum_{m_1,m_2=1}^{2n}\int_{-\infty}^\infty dt\int_0^1ds_1\,p_{\alpha,z}(t)h_{i,m_1}S_{m_1,m_2}(s_1)\bp{\bp{\sum_{\ell=1}^{2n}T_{m_2,\ell}(t)\hat{r}^{c_\mu}_\ell}+\mu_{m_2}-r_{m_2}},\right. \notag\\
    &\hspace{2.5cm}\left.-\frac{1}{2}\sum_{n_1,n_2,k_1,k_2=1}^{2n}G_{n_1,n_2}\int_0^1ds_2\,h_{j,k_1}S_{k_1,k_2}(s_2)\bp{\hat{r}^{c_\mu}_{n_1}i\Omega_{n_2,k_2}+\hat{r}^{c_\mu}_{n_2}i\Omega_{n_1,k_2}}\right]\\
    &=\frac{1}{2}\sum_{\substack{m_1,m_2,\ell \\ n_1,n_2,k_1,k_2}=1}^{2n}G_{n_1,n_2}\int_{-\infty}^\infty dt\int_0^1ds_1\int_0^1ds_2\,p_{\alpha,z}(t)h_{i,m_1}S_{m_1,m_2}(s_1)T_{m_2,\ell}(t)\times \notag\\
    &\hspace{6.5cm}h_{j,k_1}S_{k_1,k_2}(s_2)\bb{\hat{r}^{c_\mu}_\ell,\hat{r}^{c_\mu}_{n_1}i\Omega_{n_2,k_2}+\hat{r}^{c_\mu}_{n_2}i\Omega_{n_1,k_2}}\\
    &=\frac{1}{2}\sum_{\substack{m_1,m_2,\ell \\ n_1,n_2,k_1,k_2}=1}^{2n}G_{n_1,n_2}\int_{-\infty}^\infty dt\int_0^1ds_1\int_0^1ds_2\,p_{\alpha,z}(t)h_{i,m_1}S_{m_1,m_2}(s_1)T_{m_2,\ell}(t)\times \notag\\
    &\hspace{6.5cm}h_{j,k_1}S_{k_1,k_2}(s_2)\bp{i\Omega_{n_2,k_2}i\Omega_{\ell,n_1}+i\Omega_{n_1,k_2}i\Omega_{\ell,n_2}}\\
    &=-\frac{1}{2}\int_{-\infty}^\infty dt\int_0^1ds_1\int_0^1ds_2\,p_{\alpha,z}(t)\left(\bb{HS(s_1)T(t)\Omega G\Omega S^T(s_2)H^T}_{i,j}\right. \notag\\
    &\hspace{6cm}\left.+\bb{HS(s_1)T(t)\Omega G^T\Omega S^T(s_2) H^T}_{i,j}\right)\\
    &=-\int_{-\infty}^\infty dt\int_0^1ds_1\int_0^1ds_2\,p_{\alpha,z}(t)\bb{HS(s_1)T(t)\Omega G\Omega S^T(s_2)H^T}_{i,j} \label{eq:alphaz_meanvec_intermediate1}\\
    &=\int_{-\infty}^\infty dt\int_0^1ds_1\int_0^1ds_2\,p_{\alpha,z}(t)\bb{HS(s_1)T(t)\Omega G\bp{HS(s_2)\Omega}^T}_{i,j},\label{eq:alphaz_meanvec_intermediate2}
\end{align}
where we used the fact that $G=G^T$ in~\eqref{eq:alphaz_meanvec_intermediate1} and $\Omega^T=-\Omega$ in~\eqref{eq:alphaz_meanvec_intermediate2}. Taking an expectation value of~\eqref{eq:alphaz_meanvec_intermediate1} with respect to $\rho$ merely introduces a factor of $\Tr[\rho]=1$, because everything inside of the integral in~\eqref{eq:alphaz_meanvec_intermediate1} can be factored out of the trace. With this, we conclude the proof of~\eqref{eq:final_info_mtx_1}.

Now, we turn our focus to $\left[\Phi_{p_{\alpha,z}}(\Psi_\phi(\pd_{(j_1,k_1)}\hat{H}(\phi))),\bb{\hat{G},\Psi_\phi(\pd_{(j_2,k_2)}\hat{H}(\phi))}\right]$. Consider the following:
\begin{align}
    &\bb{\Phi_{p_{\alpha,z}}(\Psi_\phi(\pd_{(j_1,k_1)}\hat{H}(\phi))),\bb{\hat{G},\Psi_\phi(\pd_{(j_2,k_2)}\hat{H}(\phi))}} \notag\\
    &=\biggl[\frac{1}{4}\sum_{m_1,m_2=1}^{2n}\int_{-\infty}^\infty dt\int_0^1ds_1\,p_{\alpha,z}(t)S_{j_1,m_1}(s_1)S_{k_1,m_2}(s_1)\left(\sum_{\ell_1,\ell_2=1}^{2n}T_{m_1,\ell_1}(t)T_{m_2,\ell_2}(t)\bc{\hat{r}^{c_\mu}_{\ell_1},\hat{r}^{c_\mu}_{\ell_2}}\right. \notag\\
    &\hspace{2.2cm}\left.+2\sum_{\ell_1=1}^{2n}T_{m_1,\ell_1}(t)\hat{r}^{c_\mu}_{\ell_1}(\mu-r)_{m_2}+2\sum_{\ell_2=1}^{2n}T_{m_2,\ell_2}(t)\hat{r}^{c_\mu}_{\ell_2}(\mu-r)_{m_1}+2(\mu-r)_{m_1}(\mu-r)_{m_2}\right), \notag\\
    &\hspace{2.9cm}\frac{1}{8}\sum_{i_1,i_2=1}^{2n}\sum_{n_1,n_2=1}^{2n}G_{n_1,n_2}\int_0^1ds_2\,S_{j_2,i_1}(s_2)S_{k_2,i_2}(s_2)\bigg(i\Omega_{n_1,i_2}\bc{\hat{r}^{c_r}_{i_1},\hat{r}^{c_\mu}_{n_2}}  \notag \\
    &\hspace{4.4cm}+i\Omega_{n_2,i_2}\bc{\hat{r}^{c_r}_{i_1},\hat{r}^{c_\mu}_{n_1}}+i\Omega_{n_1,i_1}\bc{\hat{r}^{c_\mu}_{n_2},\hat{r}^{c_r}_{i_2}}+i\Omega_{n_2,i_1}\bc{\hat{r}^{c_\mu}_{n_1},\hat{r}^{c_r}_{i_2}}\bigg)   \biggl]\\
    &=\frac{1}{32}\sum_{\substack{m_1,m_2,i_1,\\i_2,n_1,n_2}=1}^{2n}\int_{-\infty}^\infty dt\int_0^1ds_1\int_0^1ds_2\,p_{\alpha,z}(t)G_{n_1,n_2}S_{j_1,m_1}(s_1)S_{k_1,m_2}(s_1)S_{j_2,i_1}(s_2)S_{k_2,i_2}(s_2)\times \notag\\
    &\hspace{0.5cm}\left(\sum_{\ell_1,\ell_2=1}^{2n}T_{m_1,\ell_1}(t)T_{m_2,\ell_2}(t)\left[\bc{\hat{r}^{c_\mu}_{\ell_1},\hat{r}^{c_\mu}_{\ell_2}},i\Omega_{n_1,i_2}\bc{\hat{r}^{c_r}_{i_1},\hat{r}^{c_\mu}_{n_2}}+i\Omega_{n_2,i_2}\bc{\hat{r}^{c_r}_{i_1},\hat{r}^{c_\mu}_{n_1}}\right.\right. \notag\\
    &\hspace{6.2cm}\left.+i\Omega_{n_1,i_1}\bc{\hat{r}^{c_\mu}_{n_2},\hat{r}^{c_r}_{i_2}}+i\Omega_{n_2,i_1}\bc{\hat{r}^{c_\mu}_{n_1},\hat{r}^{c_r}_{i_2}}\right] \notag\\
    &\hspace{0.5cm}+2\sum_{\ell_2=1}^{2n}T_{m_2,\ell_2}(t)(\mu-r)_{m_1}\left[\hat{r}^{c_\mu}_{\ell_2},i\Omega_{n_1,i_2}\bc{\hat{r}^{c_r}_{i_1},\hat{r}^{c_\mu}_{n_2}}+i\Omega_{n_2,i_2}\bc{\hat{r}^{c_r}_{i_1},\hat{r}^{c_\mu}_{n_1}}\right. \notag\\
    &\hspace{5.4cm}\left.+i\Omega_{n_1,i_1}\bc{\hat{r}^{c_\mu}_{n_2},\hat{r}^{c_r}_{i_2}}+i\Omega_{n_2,i_1}\bc{\hat{r}^{c_\mu}_{n_1},\hat{r}^{c_r}_{i_2}}\right] \notag\\
    &\hspace{0.5cm}+2\sum_{\ell_1=1}^{2n}T_{m_1,\ell_1}(t)(\mu-r)_{m_2}\left[\hat{r}^{c_\mu}_{\ell_1},i\Omega_{n_1,i_2}\bc{\hat{r}^{c_r}_{i_1},\hat{r}^{c_\mu}_{n_2}}+i\Omega_{n_2,i_2}\bc{\hat{r}^{c_r}_{i_1},\hat{r}^{c_\mu}_{n_1}}\right. \notag\\
    &\hspace{5.4cm}\left.+i\Omega_{n_1,i_1}\bc{\hat{r}^{c_\mu}_{n_2},\hat{r}^{c_r}_{i_2}}+i\Omega_{n_2,i_1}\bc{\hat{r}^{c_\mu}_{n_1},\hat{r}^{c_r}_{i_2}}\right]\biggr).
    \label{intres:info_mtx_2hamiltonian}
\end{align}
Let us simplify $\bb{\bc{\hat{r}^{c_a}_i,\hat{r}^{c_b}_j},\bc{\hat{r}^{c_c}_k,\hat{r}^{c_d}_l}}$:
\begin{align}
    \bb{\bc{\hat{r}^{c_a}_i,\hat{r}^{c_b}_j},\bc{\hat{r}^{c_c}_k,\hat{r}^{c_d}_l}}&=\bb{\hat{r}^{c_a}_i\hat{r}^{c_b}_j,\bc{\hat{r}^{c_c}_k,\hat{r}^{c_d}_l}}+\bb{\hat{r}^{c_b}_j\hat{r}^{c_a}_i,\bc{\hat{r}^{c_c}_k,\hat{r}^{c_d}_l}}\\
    & =i\Omega_{i,l}\bc{\hat{r}^{c_c}_k,\hat{r}^{c_b}_j}+i\Omega_{j,l}\bc{\hat{r}^{c_c}_k,\hat{r}^{c_a}_i}+i\Omega_{i,k}\bc{\hat{r}^{c_b}_j,\hat{r}^{c_d}_l}+i\Omega_{j,k}\bc{\hat{r}^{c_a}_i,\hat{r}^{c_d}_l} \notag\\
    &\qquad +i\Omega_{j,l}\bc{\hat{r}^{c_c}_k,\hat{r}^{c_a}_i}+i\Omega_{i,l}\bc{\hat{r}^{c_c}_k,\hat{r}^{c_b}_j}+i\Omega_{j,k}\bc{\hat{r}^{c_a}_i,\hat{r}^{c_d}_l}+i\Omega_{i,k}\bc{\hat{r}^{c_b}_j,\hat{r}^{c_d}_l}\\
    & =2\bb{\hat{r}^{c_a}_i\hat{r}^{c_b}_j,\bc{\hat{r}^{c_c}_k,\hat{r}^{c_d}_l}},
\end{align}
where \eqref{res:commutator_2quad_anticomm_2quad} was employed. Now, we will consider just the first commutator in \eqref{intres:info_mtx_2hamiltonian}:
\begin{align}
    \bigg[\bc{\hat{r}^{c_\mu}_{\ell_1},\hat{r}^{c_\mu}_{\ell_2}}&,i\Omega_{n_1,i_2}\bc{\hat{r}^{c_r}_{i_1},\hat{r}^{c_\mu}_{n_2}}+i\Omega_{n_2,i_2}\bc{\hat{r}^{c_r}_{i_1},\hat{r}^{c_\mu}_{n_1}}+i\Omega_{n_1,i_1}\bc{\hat{r}^{c_\mu}_{n_2},\hat{r}^{c_r}_{i_2}}+i\Omega_{n_2,i_1}\bc{\hat{r}^{c_\mu}_{n_1},\hat{r}^{c_r}_{i_2}}\bigg] \notag\\
    =2i\biggl({}&\Omega_{n_1,i_2}\bb{\hat{r}^{c_\mu}_{\ell_1}\hat{r}^{c_\mu}_{\ell_2},\bc{\hat{r}^{c_r}_{i_1},\hat{r}^{c_\mu}_{n_2}}}+\Omega_{n_2,i_2}\bb{\hat{r}^{c_\mu}_{\ell_1}\hat{r}^{c_\mu}_{\ell_2},\bc{\hat{r}^{c_r}_{i_1},\hat{r}^{c_\mu}_{n_1}}} \notag\\
    &+\Omega_{n_1,i_1}\bb{\hat{r}^{c_\mu}_{\ell_1}\hat{r}^{c_\mu}_{\ell_2},\bc{\hat{r}^{c_\mu}_{n_2},\hat{r}^{c_r}_{i_2}}}+\Omega_{n_2,i_1}\bb{\hat{r}^{c_\mu}_{\ell_1}\hat{r}^{c_\mu}_{\ell_2},\bc{\hat{r}^{c_\mu}_{n_1},\hat{r}^{c_r}_{i_2}}}\biggl)\\
    =-2\biggl({}&\Omega_{n_1,i_2}\bp{\Omega_{\ell_1,i_1}\bc{\hat{r}^{c_\mu}_{\ell_2},\hat{r}^{c_\mu}_{n_2}}+\Omega_{\ell_2,i_1}\bc{\hat{r}^{c_\mu}_{\ell_1},\hat{r}^{c_\mu}_{n_2}}+\Omega_{\ell_1,n_2}\bc{\hat{r}^{c_\mu}_{\ell_2},\hat{r}^{c_r}_{i_1}}+\Omega_{\ell_2,n_2}\bc{\hat{r}^{c_\mu}_{\ell_1},\hat{r}^{c_r}_{i_1}}} \notag\\
    &+\Omega_{n_2,i_2}\bp{\Omega_{\ell_1,i_1}\bc{\hat{r}^{c_\mu}_{\ell_2},\hat{r}^{c_\mu}_{n_1}}+\Omega_{\ell_2,i_1}\bc{\hat{r}^{c_\mu}_{\ell_1},\hat{r}^{c_\mu}_{n_1}}+\Omega_{\ell_1,n_1}\bc{\hat{r}^{c_\mu}_{\ell_2},\hat{r}^{c_r}_{i_1}}+\Omega_{\ell_2,n_1}\bc{\hat{r}^{c_\mu}_{\ell_1},\hat{r}^{c_r}_{i_1}}} \notag\\
    &+\Omega_{n_1,i_1}\bp{\Omega_{\ell_1,i_2}\bc{\hat{r}^{c_\mu}_{\ell_2},\hat{r}^{c_\mu}_{n_2}}+\Omega_{\ell_2,i_2}\bc{\hat{r}^{c_\mu}_{\ell_1},\hat{r}^{c_\mu}_{n_2}}+\Omega_{\ell_1,n_2}\bc{\hat{r}^{c_\mu}_{\ell_2},\hat{r}^{c_r}_{i_2}}+\Omega_{\ell_2,n_2}\bc{\hat{r}^{c_\mu}_{\ell_1},\hat{r}^{c_r}_{i_2}}} \notag\\
    &+\Omega_{n_2,i_1}\bp{\Omega_{\ell_1,i_2}\bc{\hat{r}^{c_\mu}_{\ell_2},\hat{r}^{c_\mu}_{n_1}}+\Omega_{\ell_2,i_2}\bc{\hat{r}^{c_\mu}_{\ell_1},\hat{r}^{c_\mu}_{n_1}}+\Omega_{\ell_1,n_1}\bc{\hat{r}^{c_\mu}_{\ell_2},\hat{r}^{c_r}_{i_2}}+\Omega_{\ell_2,n_1}\bc{\hat{r}^{c_\mu}_{\ell_1},\hat{r}^{c_r}_{i_2}}}\biggl).
    \label{intres:comm_anticomm_sum_omega_anticomms}
\end{align}
Note that we can write
\begin{align}
    \bc{\hat{r}^{c_\mu}_{i},\hat{r}^{c_r}_{j}}&=\bc{\hat{r}^{c_\mu}_{i},\hat{r}^{c_\mu}_{j}+\mu_j-r_j}\\
    &=\bc{\hat{r}^{c_\mu}_{i},\hat{r}^{c_\mu}_{j}}+2(\mu_j-r_j)\hat{r}^{c_\mu}_{i}. \label{eq:c_mu_relation}
\end{align}
This allows us to write \eqref{intres:comm_anticomm_sum_omega_anticomms} as
\begin{align}
    \bigg[\bc{\hat{r}^{c_\mu}_{\ell_1},\hat{r}^{c_\mu}_{\ell_2}}&,i\Omega_{n_1,i_2}\bc{\hat{r}^{c_r}_{i_1},\hat{r}^{c_\mu}_{n_2}}+i\Omega_{n_2,i_2}\bc{\hat{r}^{c_r}_{i_1},\hat{r}^{c_\mu}_{n_1}}+i\Omega_{n_1,i_1}\bc{\hat{r}^{c_\mu}_{n_2},\hat{r}^{c_r}_{i_2}}+i\Omega_{n_2,i_1}\bc{\hat{r}^{c_\mu}_{n_1},\hat{r}^{c_r}_{i_2}}\bigg] \notag\\
    =-2\biggl(&\Omega_{n_1,i_2}\biggl(\Omega_{\ell_1,i_1}\bc{\hat{r}^{c_\mu}_{\ell_2},\hat{r}^{c_\mu}_{n_2}}+\Omega_{\ell_2,i_1}\bc{\hat{r}^{c_\mu}_{\ell_1},\hat{r}^{c_\mu}_{n_2}}+\Omega_{\ell_1,n_2}\bp{\bc{\hat{r}^{c_\mu}_{\ell_2},\hat{r}^{c_\mu}_{i_1}}+2\bp{\mu_{i_1}-r_{i_1}}\hat{r}^{c_\mu}_{\ell_2}} \notag\\
    &+\Omega_{\ell_2,n_2}\bp{\bc{\hat{r}^{c_\mu}_{\ell_1},\hat{r}^{c_\mu}_{i_1}}+2\bp{\mu_{i_1}-r_{i_1}}\hat{r}^{c_\mu}_{\ell_1}}\biggr) \notag\\
    &+\Omega_{n_2,i_2}\bigg(\Omega_{\ell_1,i_1}\bc{\hat{r}^{c_\mu}_{\ell_2},\hat{r}^{c_\mu}_{n_1}}+\Omega_{\ell_2,i_1}\bc{\hat{r}^{c_\mu}_{\ell_1},\hat{r}^{c_\mu}_{n_1}}+\Omega_{\ell_1,n_1}\bp{\bc{\hat{r}^{c_\mu}_{\ell_2},\hat{r}^{c_\mu}_{i_1}}+2\bp{\mu_{i_1}-r_{i_1}}\hat{r}^{c_\mu}_{\ell_2}} \notag\\
    &+\Omega_{\ell_2,n_1}\bp{\bc{\hat{r}^{c_\mu}_{\ell_1},\hat{r}^{c_\mu}_{i_1}}+2\bp{\mu_{i_1}-r_{i_1}}\hat{r}^{c_\mu}_{\ell_1}}\biggr) \notag\\
    &+\Omega_{n_1,i_1}\biggl(\Omega_{\ell_1,i_2}\bc{\hat{r}^{c_\mu}_{\ell_2},\hat{r}^{c_\mu}_{n_2}}+\Omega_{\ell_2,i_2}\bc{\hat{r}^{c_\mu}_{\ell_1},\hat{r}^{c_\mu}_{n_2}}+\Omega_{\ell_1,n_2}\bp{\bc{\hat{r}^{c_\mu}_{\ell_2},\hat{r}^{c_\mu}_{i_2}}+2\bp{\mu_{i_2}-r_{i_2}}\hat{r}^{c_\mu}_{\ell_2}} \notag\\
    &+\Omega_{\ell_2,n_2}\bp{\bc{\hat{r}^{c_\mu}_{\ell_1},\hat{r}^{c_\mu}_{i_2}}+2\bp{\mu_{i_2}-r_{i_2}}\hat{r}^{c_\mu}_{\ell_1}}\biggr) \notag\\
    &+\Omega_{n_2,i_1}\biggl(\Omega_{\ell_1,i_2}\bc{\hat{r}^{c_\mu}_{\ell_2},\hat{r}^{c_\mu}_{n_1}}+\Omega_{\ell_2,i_2}\bc{\hat{r}^{c_\mu}_{\ell_1},\hat{r}^{c_\mu}_{n_1}}+\Omega_{\ell_1,n_1}\bp{\bc{\hat{r}^{c_\mu}_{\ell_2},\hat{r}^{c_\mu}_{i_2}}+2\bp{\mu_{i_2}-r_{i_2}}\hat{r}^{c_\mu}_{\ell_2}} \notag\\
    &+\Omega_{\ell_2,n_1}\bp{\bc{\hat{r}^{c_\mu}_{\ell_1},\hat{r}^{c_\mu}_{i_2}}+2\bp{\mu_{i_2}-r_{i_2}}\hat{r}^{c_\mu}_{\ell_1}}\biggr)\biggr).
    \label{intres:comm_anticomm_sum_omega_anticomms2}
\end{align}
Recall that the entries of the covariance matrix $V$ of $\rho$ are 
\begin{align}
    V_{j,k}&\coloneq\frac{1}{2}\ba{\bc{\hat{r}^{c_{\mu}}_j,\hat{r}^{c_{\mu}}_k}}_{\rho}.
\end{align}
With this and the fact that $\ba{\hat{r}^{c_\mu}}_\rho=0$, we find that
\begin{align}
    &\left\langle\,\bb{\bc{\hat{r}^{c_\mu}_{\ell_1},\hat{r}^{c_\mu}_{\ell_2}},i\Omega_{n_1,i_2}\bc{\hat{r}^{c_r}_{i_1},\hat{r}^{c_\mu}_{n_2}}+i\Omega_{n_2,i_2}\bc{\hat{r}^{c_r}_{i_1},\hat{r}^{c_\mu}_{n_1}}+i\Omega_{n_1,i_1}\bc{\hat{r}^{c_\mu}_{n_2},\hat{r}^{c_r}_{i_2}}+i\Omega_{n_2,i_1}\bc{\hat{r}^{c_\mu}_{n_1},\hat{r}^{c_r}_{i_2}}}\right\rangle_\rho \notag\\
    &=-4\biggl( \Omega_{n_1,i_2}\left(\Omega_{\ell_1,i_1}V_{\ell_2,n_2}+\Omega_{\ell_2,i_1}V_{\ell_1,n_2}+\Omega_{\ell_1,n_2}V_{\ell_2,i_1}+\Omega_{\ell_2,n_2}V_{\ell_1,i_1}\right)  \notag\\
    &\qquad +\Omega_{n_2,i_2}\left(\Omega_{\ell_1,i_1}V_{\ell_2,n_1}+\Omega_{\ell_2,i_1}V_{\ell_1,n_1}+\Omega_{\ell_1,n_1}V_{\ell_2,i_1}+\Omega_{\ell_2,n_1}V_{\ell_1,i_1}\right) \notag\\
    &\qquad +\Omega_{n_1,i_1}\left(\Omega_{\ell_1,i_2}V_{\ell_2,n_2}+\Omega_{\ell_2,i_2}V_{\ell_1,n_2}+\Omega_{\ell_1,n_2}V_{\ell_2,i_2}+\Omega_{\ell_2,n_2}V_{\ell_1,i_2}\right) \notag\\
    &\qquad +\Omega_{n_2,i_1}\left(\Omega_{\ell_1,i_2}V_{\ell_2,n_1}+\Omega_{\ell_2,i_2}V_{\ell_1,n_1}+\Omega_{\ell_1,n_1}V_{\ell_2,i_2}+\Omega_{\ell_2,n_1}V_{\ell_1,i_2}\right)\biggr).\label{eq:big_expec1}
\end{align}
We will now evaluate the second commutator in~\eqref{intres:info_mtx_2hamiltonian}:
\begin{align}
    &\bb{\hat{r}^{c_\mu}_{\ell_2},i\Omega_{n_1,i_2}\bc{\hat{r}^{c_r}_{i_1},\hat{r}^{c_\mu}_{n_2}}+i\Omega_{n_2,i_2}\bc{\hat{r}^{c_r}_{i_1},\hat{r}^{c_\mu}_{n_1}}+i\Omega_{n_1,i_1}\bc{\hat{r}^{c_\mu}_{n_2},\hat{r}^{c_r}_{i_2}}+i\Omega_{n_2,i_1}\bc{\hat{r}^{c_\mu}_{n_1},\hat{r}^{c_r}_{i_2}}} \notag\\
    &=2\Omega_{n_1,i_2}\bp{\Omega_{n_2,\ell_2}\hat{r}^{c_r}_{i_1}+\Omega_{i_1,\ell_2}\hat{r}^{c_\mu}_{n_2}}+2\Omega_{n_2,i_2}\bp{\Omega_{n_1,\ell_2}\hat{r}^{c_r}_{i_1}+\Omega_{i_1,\ell_2}\hat{r}^{c_\mu}_{n_1}}+2\Omega_{n_1,i_1}\bp{\Omega_{i_2,\ell_2}\hat{r}^{c_\mu}_{n_2}+\Omega_{n_2,\ell_2}\hat{r}^{c_r}_{i_2}} \notag\\
    &\hspace{10cm}+2\Omega_{n_2,i_1}\bp{\Omega_{i_2,\ell_2}\hat{r}^{c_\mu}_{n_1}+\Omega_{n_1,\ell_2}\hat{r}^{c_r}_{i_2}}.
\end{align}
Noting that $\ba{\hat{r}^{c_\mu}}_\rho=0$ and $\ba{\hat{r}^{c_r}}_\rho=\mu-r$, we find that
\begin{align}
    &\left\langle\,\bb{\hat{r}^{c_\mu}_{\ell_2},i\Omega_{n_1,i_2}\bc{\hat{r}^{c_r}_{i_1},\hat{r}^{c_\mu}_{n_2}}+i\Omega_{n_2,i_2}\bc{\hat{r}^{c_r}_{i_1},\hat{r}^{c_\mu}_{n_1}}+i\Omega_{n_1,i_1}\bc{\hat{r}^{c_\mu}_{n_2},\hat{r}^{c_r}_{i_2}}+i\Omega_{n_2,i_1}\bc{\hat{r}^{c_\mu}_{n_1},\hat{r}^{c_r}_{i_2}}}\right\rangle_\rho \notag\\
    &=2\Omega_{n_1,i_2}\Omega_{n_2,\ell_2}(\mu-r)_{i_1}+2\Omega_{n_2,i_2}\Omega_{n_1,\ell_2}(\mu-r)_{i_1}+2\Omega_{n_1,i_1}\Omega_{n_2,\ell_2}(\mu-r)_{i_2}+2\Omega_{n_2,i_1}\Omega_{n_1,\ell_2}(\mu-r)_{i_2}.\label{eq:big_expec2}
\end{align}
We can now employ~\eqref{eq:big_expec1} and~\eqref{eq:big_expec2} along with~\eqref{intres:info_mtx_2hamiltonian} to conclude that
\begin{align}
    &\left\langle\,\bb{\Phi_{p_{\alpha,z}}(\Psi_\phi(\pd_{(j_1,k_1)}\hat{H}(\phi))),\bb{\hat{G},\Psi_\phi(\pd_{(j_2,k_2)}\hat{H}(\phi))}}\right\rangle_\rho \notag\\
    &=-\frac{1}{8}\int_{-\infty}^\infty dt\int_0^1ds_1\int_0^1ds_2\,p_{\alpha,z}(t)\sum_{\substack{m_1,m_2,i_1,\\i_2,n_1,n_2}=1}^{2n}G_{n_1,n_2}S_{j_1,m_1}(s_1)S_{k_1,m_2}(s_1)S_{j_2,i_1}(s_2)S_{k_2,i_2}(s_2)\times \notag\\
    &\biggl(\sum_{\ell_1,\ell_2=1}^{2n}T_{m_1,\ell_1}(t)T_{m_2,\ell_2}(t)\biggl( \Omega_{n_1,i_2}\left(\Omega_{\ell_1,i_1}V_{\ell_2,n_2}+\Omega_{\ell_2,i_1}V_{\ell_1,n_2}+\Omega_{\ell_1,n_2}V_{\ell_2,i_1}+\Omega_{\ell_2,n_2}V_{\ell_1,i_1}\right)  \notag\\
    &\hspace{3.3cm}+\Omega_{n_2,i_2}\left(\Omega_{\ell_1,i_1}V_{\ell_2,n_1}+\Omega_{\ell_2,i_1}V_{\ell_1,n_1}+\Omega_{\ell_1,n_1}V_{\ell_2,i_1}+\Omega_{\ell_2,n_1}V_{\ell_1,i_1}\right) \notag\\
    &\hspace{3.3cm}+\Omega_{n_1,i_1}\left(\Omega_{\ell_1,i_2}V_{\ell_2,n_2}+\Omega_{\ell_2,i_2}V_{\ell_1,n_2}+\Omega_{\ell_1,n_2}V_{\ell_2,i_2}+\Omega_{\ell_2,n_2}V_{\ell_1,i_2}\right) \notag\\
    &\hspace{3.3cm}+\Omega_{n_2,i_1}\left(\Omega_{\ell_1,i_2}V_{\ell_2,n_1}+\Omega_{\ell_2,i_2}V_{\ell_1,n_1}+\Omega_{\ell_1,n_1}V_{\ell_2,i_2}+\Omega_{\ell_2,n_1}V_{\ell_1,i_2}\right) \biggr)\notag\\
    &-\sum_{\ell_2=1}^{2n}T_{m_2,\ell_2}(t)(\mu-r)_{m_1}\left(\Omega_{n_1,i_2}\Omega_{n_2,\ell_2}(\mu-r)_{i_1}+\Omega_{n_2,i_2}\Omega_{n_1,\ell_2}(\mu-r)_{i_1}\right. \notag\\
    &\hspace{4cm}\left.+\Omega_{n_1,i_1}\Omega_{n_2,\ell_2}(\mu-r)_{i_2}+\Omega_{n_2,i_1}\Omega_{n_1,\ell_2}(\mu-r)_{i_2}\right) \notag\\
    &-\sum_{\ell_1=1}^{2n}T_{m_1,\ell_1}(t)(\mu-r)_{m_2}\left(\Omega_{n_1,i_2}\Omega_{n_2,\ell_1}(\mu-r)_{i_1}+\Omega_{n_2,i_2}\Omega_{n_1,\ell_1}(\mu-r)_{i_1}\right. \notag\\
    &\hspace{4cm}\left.+\Omega_{n_1,i_1}\Omega_{n_2,\ell_1}(\mu-r)_{i_2}+\Omega_{n_2,i_1}\Omega_{n_1,\ell_1}(\mu-r)_{i_2}\right)\biggr).\label{eq:big_sum_two_ham}
\end{align}
Let $\widetilde{\xi}_{j_1,k_1,j_2,k_2}(s_1,s_2,t)$ denote the outer sum in~\eqref{eq:big_sum_two_ham}. It follows that
\begin{align}
    &\widetilde{\xi}_{j_1,k_1,j_2,k_2}(s_1,s_2,t) \notag\\
    &=\bb{S(s_1)T\Omega S^T(s_2)}_{j_1,j_2}\bb{S(s_1)TVG\Omega S^T(s_2)}_{k_1,k_2}+\bb{S(s_1)TVG\Omega S^T(s_2)}_{j_1,k_2}\bb{S(s_1)T\Omega S^T(s_2)}_{k_1,j_2} \notag\\
    &+\bb{S(s_1)T\Omega G\Omega S^T(s_2)}_{j_1,k_2}\bb{S(s_1)TV S^T(s_2)}_{k_1,j_2}+\bb{S(s_1)TV S^T(s_2)}_{j_1,j_2}\bb{S(s_1)T\Omega G\Omega S^T(s_2)}_{k_1,k_2} \notag\\
    &+\bb{S(s_1)T\Omega S^T(s_2)}_{j_1,j_2}\bb{S(s_1)TVG\Omega S^T(s_2)}_{k_1,k_2}+\bb{S(s_1)TVG\Omega S^T(s_2)}_{j_1,k_2}\bb{S(s_1)T\Omega S^T(s_2)}_{k_1,j_2} \notag\\
    &+\bb{S(s_1)T\Omega G\Omega S^T(s_2)}_{j_1,k_2}\bb{S(s_1)TV S^T(s_2)}_{k_1,j_2}+\bb{S(s_1)TV S^T(s_2)}_{j_1,j_2}\bb{S(s_1)T\Omega G\Omega S^T(s_2)}_{k_1,k_2} \notag\\
    &+\bb{S(s_1)T\Omega S^T(s_2)}_{j_1,k_2}\bb{S(s_1)TVG\Omega S^T(s_2)}_{k_1,j_2}+\bb{S(s_1)TV G \Omega S^T(s_2)}_{j_1,j_2}\bb{S(s_1)T\Omega S^T(s_2)}_{k_1,k_2} \notag\\
    &+\bb{S(s_1)T\Omega G\Omega S^T(s_2)}_{j_1,j_2}\bb{S(s_1)TV S^T(s_2)}_{k_1,k_2}+\bb{S(s_1)TV S^T(s_2)}_{j_1,k_2}\bb{S(s_1)T\Omega G\Omega S^T(s_2)}_{k_1,j_2} \notag\\
    &+\bb{S(s_1)T\Omega S^T(s_2)}_{j_1,k_2}\bb{S(s_1)TV G\Omega S^T(s_2)}_{k_1,j_2}+\bb{S(s_1)TV G\Omega S^T(s_2)}_{j_1,j_2}\bb{S(s_1)T\Omega S^T(s_2)}_{k_1,k_2} \notag\\
    &+\bb{S(s_1)T\Omega G\Omega S^T(s_2)}_{j_1,j_2}\bb{S(s_1)TV S^T(s_2)}_{k_1,k_2}+\bb{S(s_1)TV S^T(s_2)}_{j_1,k_2}\bb{S(s_1)T\Omega G\Omega S^T(s_2)}_{k_1,j_2} \notag\\
    &-\bb{S(s_1)(\mu-r)}_{j_1}\left(\bb{S(s_1)T \Omega^TG\Omega S^T(s_2)}_{k_1,k_2}\bb{S(s_2)(\mu-r)}_{j_2}\right. \notag\\
    &\hspace{6cm}+\bb{S(s_1)T \Omega^TG\Omega S^T(s_2)}_{k_1,k_2}\bb{S(s_2)(\mu-r)}_{j_2} \notag\\
    &\hspace{1cm}\left.+\bb{S(s_1)T \Omega^T G\Omega S^T(s_2)}_{k_1,j_2}\bb{S(s_2)(\mu-r)}_{k_2}+\bb{S(s_1)T \Omega^T G\Omega S^T(s_2)}_{k_1,j_2}\bb{S(s_2)(\mu-r)}_{k_2}\right) \notag\\
    &-\bb{S(s_1)(\mu-r)}_{k_1}\left(\bb{S(s_1)T \Omega^TG\Omega S^T(s_2)}_{j_1,k_2}\bb{S(s_2)(\mu-r)}_{j_2}\right. \notag\\
    &\hspace{4cm}+\bb{S(s_1)T \Omega^TG\Omega S^T(s_2)}_{j_1,k_2}\bb{S(s_2)(\mu-r)}_{j_2} \notag\\
    &\hspace{1cm}\left.+\bb{S(s_1)T \Omega^T G\Omega S^T(s_2)}_{j_1,j_2}\bb{S(s_2)(\mu-r)}_{k_2}+\bb{S(s_1)T \Omega^T G\Omega S^T(s_2)}_{j_1,j_2}\bb{S(s_2)(\mu-r)}_{k_2}\right)\\
    &=2\bb{S(s_1)T\Omega S^T(s_2)}_{j_1,j_2}\bb{S(s_1)TVG\Omega S^T(s_2)}_{k_1,k_2}+2\bb{S(s_1)T\Omega S^T(s_2)}_{j_1,k_2}\bb{S(s_1)TVG\Omega S^T(s_2)}_{k_1,j_2} \notag\\
    &+2\bb{S(s_1)T\Omega S^T(s_2)}_{k_1,j_2}\bb{S(s_1)TVG\Omega S^T(s_2)}_{j_1,k_2}+2\bb{S(s_1)T\Omega S^T(s_2)}_{k_1,k_2}\bb{S(s_1)TV G \Omega S^T(s_2)}_{j_1,j_2} \notag\\
    &+2\bb{S(s_1)T\Omega G\Omega S^T(s_2)}_{j_1,k_2}\bb{S(s_1)TV S^T(s_2)}_{k_1,j_2}+2\bb{S(s_1)T\Omega G\Omega S^T(s_2)}_{j_1,j_2}\bb{S(s_1)TV S^T(s_2)}_{k_1,k_2} \notag\\
    &+2\bb{S(s_1)T\Omega G\Omega S^T(s_2)}_{k_1,k_2}\bb{S(s_1)TV S^T(s_2)}_{j_1,j_2}+2\bb{S(s_1)T\Omega G\Omega S^T(s_2)}_{k_1,j_2}\bb{S(s_1)TV S^T(s_2)}_{j_1,k_2} \notag\\
    &-2\bb{S(s_1)(\mu-r)}_{j_1}\left(\bb{S(s_1)T \Omega^TG\Omega S^T(s_2)}_{k_1,k_2}\bb{S(s_2)(\mu-r)}_{j_2}\right. \notag\\
    &\hspace{4cm}\left.+\bb{S(s_1)T \Omega^TG\Omega S^T(s_2)}_{k_1,j_2}\bb{S(s_2)(\mu-r)}_{k_2}\right) \notag\\
    &-2\bb{S(s_1)(\mu-r)}_{k_1}\left(\bb{S(s_1)T \Omega^TG\Omega S^T(s_2)}_{j_1,k_2}\bb{S(s_2)(\mu-r)}_{j_2}\right. \notag\\
    &\hspace{4cm}\left.+\bb{S(s_1)T \Omega^T G\Omega S^T(s_2)}_{j_1,j_2}\bb{S(s_2)(\mu-r)}_{k_2}\right).\label{eq:big_xi}
\end{align}
Utilizing~\eqref{def:B1}--\eqref{def:bvec}, Eq.~\eqref{eq:big_xi} becomes
\begin{align}
    \widetilde{\xi}_{j_1,k_1,j_2,k_2}(s_1,s_2,t)&=2\left(\bb{B_1}_{j_1,j_2}\bb{B_2}_{k_1,k_2}+\bb{B_1}_{j_1,k_2}\bb{B_2}_{k_1,j_2}+\bb{B_1}_{k_1,j_2}\bb{B_2}_{j_1,k_2}+\bb{B_1}_{k_1,k_2}\bb{B_2}_{j_1,j_2}\right. \notag\\
    &\hspace{.8cm}+\bb{B_3}_{j_1,j_2}\bb{B_4}_{k_1,k_2}+\bb{B_3}_{j_1,k_2}\bb{B_4}_{k_1,j_2}+\bb{B_3}_{k_1,j_2}\bb{B_4}_{j_1,k_2}+\bb{B_3}_{k_1,k_2}\bb{B_4}_{j_1,j_2} \notag\\
    &\hspace{5cm}+[b(s_1)]_{j_1}\bp{\bb{B_3}_{k_1,k_2}\bb{b(s_2)}_{j_2}+\bb{B_3}_{k_1,j_2}\bb{b(s_2)}_{k_2}} \notag\\
    &\hspace{5cm}\left.+[b(s_1)]_{k_1}\bp{\bb{B_3}_{j_1,k_2}\bb{b(s_2)}_{j_2}+\bb{B_3}_{j_1,j_2}\bb{b(s_2)}_{k_2}}\right).
\end{align}
After substituting this into~\eqref{eq:big_sum_two_ham}, we conclude the proof of~\eqref{eq:final_info_mtx_2}. 

Now, we determine the cross terms of the $\alpha$-$z$ information matrix. Namely, we will find expressions for $[I_{\alpha, z}(r, H)]_{(j,k),i}$ and $[I_{\alpha, z}(r, H)]_{i,(j,k)}$.
We will begin by evaluating $\Big[\Phi_{p_{\alpha,z}}(\Psi_{r,H}(\pd_{(j,k)}\hat{H}(\phi))), [\hat{G}, \Psi_{r,H}(\pd_i\hat{H}(\phi))]\Big]$. Consider that
\begin{multline}\label{eq:com2}
        \Big[\Phi_{p_{\alpha,z}}(\Psi_{r,H}(\pd_{(j,k)}\hat{H}(\phi))), [\hat{G}, \Psi_{r,H}(\pd_i\hat{H}(\phi))]\Big] = -\frac{1}{8} \sum^{2n}_{\substack{a_1,a_2,n_1,\\n_2, m_1, m_2}=1} \int^{\infty}_{-\infty}dt\int_0^1 ds_1\int_0^1 ds_2 \: p_{\alpha,z}(t)\times\\ 
        S_{j, a_1}(s_1)S_{k,a_2}(s_1)G_{n_1,n_2}h_{i,m_1}S_{m_1,m_2}(s_2)\left[\left(\sum_{\ell_1,\ell_2=1}^{2n}T_{a_1,\ell_1}(t)T_{a_2,\ell_2}(t)\{\hat{r}_{\ell_1}^{c_\mu}, \hat{r}_{\ell_2}^{c_\mu}\}+2\sum_{\ell_1=1}^{2n}(\mu-r)_{a_2}T_{a_1,\ell_1}(t)\hat{r}_{\ell_1}^{c_\mu}\right.\right.\\
        \left.\left.+ 2\sum_{\ell_2=1}^{2n}(\mu-r)_{a_1}T_{a_2,\ell_2}(t)\hat{r}_{\ell_2}^{c_\mu}+2(\mu-r)_{a_1}(\mu-r)_{a_2}\right), \bp{\hat{r}_{n_1}^{c_\mu}i\Omega_{n_2,m_2}+ \hat{r}_{n_2}^{c_\mu}i\Omega_{n_1,m_2}}\right].
\end{multline}
The following relation will be useful:
\begin{align} 
        \bb{\hat{r}^{c_a}_a, \bc{\hat{r}^{c_b}_b,\hat{r}^{c_c}_c}} &= \hat{r}^{c_a}_a\hat{r}^{c_b}_b\hat{r}^{c_c}_c + \hat{r}^{c_a}_a\hat{r}^{c_c}_c\hat{r}^{c_b}_b - \hat{r}^{c_b}_b\hat{r}^{c_c}_c\hat{r}^{c_a}_a - \hat{r}^{c_c}_c\hat{r}^{c_b}_b\hat{r}^{c_a}_a \\
        &= \hat{r}^{c_a}_a\hat{r}^{c_b}_b\hat{r}^{c_c}_c + \hat{r}^{c_a}_a\hat{r}^{c_c}_c\hat{r}^{c_b}_b - \hat{r}^{c_b}_b\hat{r}^{c_a}_a\hat{r}^{c_c}_c - \hat{r}^{c_c}_c\hat{r}^{c_a}_a\hat{r}^{c_b}_b
        -i\Omega_{c,a}\hat{r}_b^{c_b}-i\Omega_{b,a}\hat{r}_c^{c_c} \\
        &= -2i\bp{\Omega_{c,a}\hat{r}_b^{c_b}+ \Omega_{b,a}\hat{r}_c^{c_c}}.\label{eq:comrel}
\end{align}
We will investigate the commutator in~\eqref{eq:com2} term by term. For a fixed $\ell_1$ and $\ell_2$, using \eqref{eq:comrel}, the first term contains
\begin{align}
        & \bb{\bc{\hat{r}_{\ell_1}^{c_\mu}, \hat{r}_{\ell_2}^{c_\mu}}, \bp{\hat{r}_{n_1}^{c_\mu}i\Omega_{n_2,m_2}+ \hat{r}_{n_2}^{c_\mu}i\Omega_{n_1,m_2}}} \notag\\
        &=
        i\Omega_{n_2,m_2}\bb{\bc{\hat{r}_{\ell_1}^{c_\mu}, \hat{r}_{\ell_2}^{c_\mu}}, \hat{r}_{n_1}^{c_\mu}} + i\Omega_{n_1,m_2}\bb{\bc{\hat{r}_{\ell_1}^{c_\mu}, \hat{r}_{\ell_2}^{c_\mu}}, \hat{r}_{n_2}^{c_\mu}} \\
        &= -2\Omega_{n_2,m_2}\bp{\Omega_{\ell_2,n_1}\hat{r}_{\ell_1}^{c_\mu}+ \Omega_{\ell_1,n_1}\hat{r}_{\ell_2}^{c_\mu}}-2\Omega_{n_1,m_2}\bp{\Omega_{\ell_2,n_2}\hat{r}_{\ell_1}^{c_\mu}+ \Omega_{\ell_1,n_2}\hat{r}_{\ell_2}^{c_\mu}} \\
        &=-2\Big(\Omega_{n_2,m_2}\Omega_{\ell_2,n_1}\hat{r}_{\ell_1}^{c_\mu} +  \Omega_{n_2,m_2}\Omega_{\ell_1,n_1}\hat{r}_{\ell_2}^{c_\mu}+ \Omega_{n_1,m_2}\Omega_{\ell_2,n_2}\hat{r}_{\ell_1}^{c_\mu} + \Omega_{n_1,m_2}\Omega_{\ell_1,n_2}\hat{r}_{\ell_2}^{c_\mu}\Big).
\end{align}
Notice that all four terms here contain quadrature operators centered by $\mu$. Because our initial state $\rho$ has mean $\mu$, all four of these terms will vanish when taking the expectation value $\langle \hat{r}^{c_\mu}\rangle_{\rho}$. Thus, the first sum in the commutator has no effect on the final expression. For a fixed $\ell_1$ or $\ell_2$, respectively, the next two terms are
\begin{align}
    2(\mu-r)_{a_2}T_{a_1,\ell_1}(t)\Big[\hat{r}_{\ell_1}^{c_\mu}, (\hat{r}_{n_1}^{c_\mu}i\Omega_{n_2,m_2}+\hat{r}_{n_2}^{c_\mu}i\Omega_{n_1,m_2})\Big] &= -2(\mu-r)_{a_2}T_{a_1,\ell_1}(t)(\Omega_{n_2,m_2}\Omega_{\ell_1,n_1}+\Omega_{n_1,m_2}\Omega_{\ell_1,n_2}),\\
    2(\mu-r)_{a_1}T_{a_2,\ell_2}(t)\Big[\hat{r}_{\ell_2}^{c_\mu}, (\hat{r}_{n_1}^{c_\mu}i\Omega_{n_2,m_2}+\hat{r}_{n_2}^{c_\mu}i\Omega_{n_1,m_2})\Big] &= -2(\mu-r)_{a_1}T_{a_2,\ell_2}(t)(\Omega_{n_2,m_2}\Omega_{\ell_2,n_1}+\Omega_{n_1,m_2}\Omega_{\ell_2,n_2}).
\end{align}
The last term in the first argument of the commutator in~\eqref{eq:com2} is just a constant, so its contribution to the commutator is zero. Thus, the final expression for the expected value of the commutator in~\eqref{eq:com2} is
\begin{multline}
-2(\mu-r)_{a_2}\sum_{\ell_1=1}^{2n}(\Omega_{n_2,m_2}\Omega_{\ell_1,n_1}+\Omega_{n_1,m_2}\Omega_{\ell_1,n_2})T_{a_1,\ell_1}(t) \\-2(\mu-r)_{a_1}\sum_{\ell_2=1}^{2n}(\Omega_{n_2,m_2}\Omega_{\ell_2,n_1}+\Omega_{n_1,m_2}\Omega_{\ell_2,n_2})T_{a_2,\ell_2}(t).
\end{multline}
Now we can combine this with the rest of \eqref{eq:com2} to get
\begin{multline}\label{eq:result2}
        [I_{\alpha, z}(r, H)]_{(j,k),i} = \frac{1}{2}  \int^{\infty}_{-\infty}dt\int_0^1 ds_1\int_0^1 ds_2 \: p_{\alpha,z}(t) \\
        \times \Big([S(s_1)T(t)\Omega G \Omega S^T(s_2)H]_{j,i}[S(s_1)(\mu-r)]_k \\+ [S(s_1)T(t)\Omega G \Omega S^T(s_2)H]_{k,i}[S(s_1)(\mu-r)]_j \Big ),
\end{multline}
completing the proof of the first part of~\eqref{eq:final_info_mtx_3}.
Now we move on to prove the second part of~\eqref{eq:final_info_mtx_3}, where the commutator is
\begin{multline}\label{eq:com}
\Big[\Phi_{p_{\alpha,z}}(\Psi_{r,H}(\pd_i\hat{H}(\phi))), [\hat{G}, \Psi_{r,H}(\pd_{(j,k)}\hat{H}(\phi))]\Big] = -\frac{1}{8}\sum^{2n}_{\substack{a_1,a_2,m_1,\\ m_2, n_1, n_2} = 1} \int_{-\infty}^{\infty}dt\int_0^1 ds_1\int_0^1 ds_2 \, \,p_{\alpha,z}(t)h_{i,a_1}
\\ \times S_{a_1,a_2}(s_1)G_{n_1,n_2}S_{j,m_1}(s_2)S_{k,m_2}(s_2)\left[\bp{\sum_{\ell=1}^{2n}T_{a_2,\ell}(t)\hat{r}_\ell^{c_\mu}}+(\mu-r)_{a_2}, i\Omega_{n_1,m_2} \{\hat{r}_{m_1}^{c_r}, \hat{r}_{n_2}^{c_\mu}\}\right. \\
+ i\Omega_{n_2,m_2} \{\hat{r}_{m_1}^{c_r}, \hat{r}_{n_1}^{c_\mu}\} +i\Omega_{n_1,m_1} \{\hat{r}_{n_2}^{c_\mu}, \hat{r}_{m_2}^{c_r}\}+i\Omega_{n_2,m_1} \{\hat{r}_{n_1}^{c_\mu}, \hat{r}_{m_2}^{c_r}\}\Big].
\end{multline}
In order to put everything in terms of $\hat{r}^{c_\mu}$, we will use \eqref{eq:c_mu_relation}, which states that 
\begin{equation}
    \bc{\hat{r}^{c_\mu}_{i},\hat{r}^{c_r}_{j}} = \bc{\hat{r}^{c_\mu}_{i},\hat{r}^{c_\mu}_{j}}+2(\mu_j-r_j)\hat{r}^{c_\mu}_{i}.
\end{equation}
The commutator is now
\begin{multline}
    \Big[\hat{r}_\ell^{c_\mu}, i\Omega_{n_1,m_2} \{\hat{r}_{m_1}^{c_r}, \hat{r}_{n_2}^{c_\mu}\} + i\Omega_{n_2,m_2} \{\hat{r}_{m_1}^{c_r}, \hat{r}_{n_1}^{c_\mu}\} +i\Omega_{n_1,m_1} \{\hat{r}_{n_2}^{c_\mu}, \hat{r}_{m_2}^{c_r}\}+i\Omega_{n_2,m_1} \{\hat{r}_{n_1}^{c_\mu}, \hat{r}_{m_2}^{c_r}\}\Big] =\\
    \qquad \qquad i\Omega_{n_1,m_2}[\hat{r}^{c_\mu}_\ell, \{\hat{r}^{c_\mu}_{n_2},\hat{r}^{c_\mu}_{m_1}\}+ 2(\mu_{m_1}-r_{m_1})\hat{r}^{c_\mu}_{n_2}]
    +i\Omega_{n_2,m_2}[\hat{r}^{c_\mu}_\ell, \{\hat{r}^{c_\mu}_{n_1},\hat{r}^{c_\mu}_{m_1}\}+ 2(\mu_{m_1}-r_{m_1})\hat{r}^{c_\mu}_{n_1}] \\
    +i\Omega_{n_1,m_1}[\hat{r}^{c_\mu}_\ell, \{\hat{r}^{c_\mu}_{n_2},\hat{r}^{c_\mu}_{m_2}\}+ 2(\mu_{m_2}-r_{m_2})\hat{r}^{c_\mu}_{n_2}]
    +i\Omega_{n_2,m_1}[\hat{r}^{c_\mu}_\ell, \{\hat{r}^{c_\mu}_{n_1},\hat{r}^{c_\mu}_{m_2}\}+ 2(\mu_{m_2}-r_{m_2})\hat{r}^{c_\mu}_{n_1}].
\end{multline}
We now use \eqref{eq:comrel} to simplify this further:
\begin{multline}
\Big[\hat{r}_\ell^{c_\mu}, i\Omega_{n_1,m_2} \{\hat{r}_{m_1}^{c_r}, \hat{r}_{n_2}^{c_\mu}\} + i\Omega_{n_2,m_2} \{\hat{r}_{m_1}^{c_r}, \hat{r}_{n_1}^{c_\mu}\} +i\Omega_{n_1,m_1} \{\hat{r}_{n_2}^{c_\mu}, \hat{r}_{m_2}^{c_r}\}+i\Omega_{n_2,m_1} \{\hat{r}_{n_1}^{c_\mu}, \hat{r}_{m_2}^{c_r}\}\Big]\\
    =i\Omega_{n_1, m_2}(-2i\Omega_{m_1,\ell}\,\hat{r}^{c_\mu}_{n_2}-2i\Omega_{n_2,\ell}\, \hat{r}^{c_\mu}_{m_1})+2i\Omega_{n_1,m_2}(\mu_{m_1}-r_{m_1})[\hat{r}^{c_\mu}_\ell,\hat{r}^{c_\mu}_{n_2}] \\
    +i\Omega_{n_2, m_2}(-2i\Omega_{m_1,\ell}\,\hat{r}^{c_\mu}_{n_1}-2i\Omega_{n_1,\ell}\, \hat{r}^{c_\mu}_{m_1})+2i\Omega_{n_2,m_2}(\mu_{m_1}-r_{m_1})[\hat{r}^{c_\mu}_\ell,\hat{r}^{c_\mu}_{n_1}]\\ 
    +i\Omega_{n_1, m_1}(-2i\Omega_{m_2,\ell}\,\hat{r}^{c_\mu}_{n_2}-2i\Omega_{n_2,\ell}\, \hat{r}^{c_\mu}_{m_2})+2i\Omega_{n_1,m_1}(\mu_{m_2}-r_{m_2})[\hat{r}^{c_\mu}_\ell,\hat{r}^{c_\mu}_{n_2}]\\
    +i\Omega_{n_2, m_1}(-2i\Omega_{m_2,\ell}\,\hat{r}^{c_\mu}_{n_1}-2i\Omega_{n_1,\ell}\, \hat{r}^{c_\mu}_{m_2})+2i\Omega_{n_2,m_1}(\mu_{m_2}-r_{m_2})[\hat{r}^{c_\mu}_\ell,\hat{r}^{c_\mu}_{n_1}].
\end{multline}
Recalling that $[\hat{r}_i^{c_\mu},\hat{r}_j^{c_\mu}] = i\Omega_{i,j}$, we find that
\begin{multline}
\Big[\hat{r}_\ell^{c_\mu}, i\Omega_{n_1,m_2} \{\hat{r}_{m_1}^{c_r}, \hat{r}_{n_2}^{c_\mu}\} + i\Omega_{n_2,m_2} \{\hat{r}_{m_1}^{c_r}, \hat{r}_{n_1}^{c_\mu}\} +i\Omega_{n_1,m_1} \{\hat{r}_{n_2}^{c_\mu}, \hat{r}_{m_2}^{c_r}\}+i\Omega_{n_2,m_1} \{\hat{r}_{n_1}^{c_\mu}, \hat{r}_{m_2}^{c_r}\}\Big]\\
    =2\Omega_{n_1, m_2}(\Omega_{m_1,\ell}\,\hat{r}^{c_\mu}_{n_2}+\Omega_{n_2,\ell}\, \hat{r}^{c_\mu}_{m_1})-2(\mu_{m_1}-r_{m_1})\Omega_{n_1,m_2}\Omega_{\ell,n_2} \\
    +2\Omega_{n_2, m_2}(\Omega_{m_1,\ell}\,\hat{r}^{c_\mu}_{n_1}+\Omega_{n_1,\ell}\, \hat{r}^{c_\mu}_{m_1})-2(\mu_{m_1}-r_{m_1})\Omega_{n_2,m_2}\Omega_{\ell,n_1}\\ 
    +2\Omega_{n_1, m_1}(\Omega_{m_2,\ell}\,\hat{r}^{c_\mu}_{n_2}+\Omega_{n_2,\ell}\, \hat{r}^{c_\mu}_{m_2})-2(\mu_{m_2}-r_{m_2})\Omega_{n_1,m_1}\Omega_{\ell,n_2}\\
    +2\Omega_{n_2, m_1}(\Omega_{m_2,\ell}\,\hat{r}^{c_\mu}_{n_1}+\Omega_{n_1,\ell}\, \hat{r}^{c_\mu}_{m_2})-2(\mu_{m_2}-r_{m_2})\Omega_{n_2,m_1}\Omega_{\ell,n_1}.
\end{multline}
Following the same reasoning as before, every term that is first order in $\hat{r}^{c_\mu}$ will vanish after taking the expectation value. That leaves us with the four scalar terms on the right. Combining this result with \eqref{eq:com}, we find that
\begin{multline}
[I_{\alpha, z}(r, H)]_{i,(j,k)} = \frac{1}{2} \int_{-\infty}^{\infty}dt\int_0^1 ds_1\int_0^1 ds_2 \: p_{\alpha,z}(t)\\
\times \Big([HS(s_1)T(t)\Omega G\Omega S^{T}(s_2)]_{i,k}[S(s_2)(\mu - r)]_j\\
+ [HS(s_1)T(t)\Omega G\Omega S^{T}(s_2)]_{i,j}[S(s_2)(\mu - r)]_k\Big ).\label{eq:cross_term1}
\end{multline}
Because the QFI matrix is symmetric, our expressions for the cross terms should be equal. We will show that this is indeed true by rewriting~\eqref{eq:result2} to match~\eqref{eq:cross_term1}.
Consider that
\begin{align}
    \bb{T(t),\Omega G}=\bb{e^{-\Omega Gt},\Omega G}=0.
\end{align}
Also, because $e^{-\Omega Gt}$ is symplectic,
\begin{align}
   e^{-\Omega Gt}\Omega=\Omega (e^{-\Omega Gt})^{-T}=\Omega(e^{\Omega Gt})^{T}.
\end{align}
Consider the following matrix element in~\eqref{eq:result2}:
\begin{align}
    \bb{S(s_1)T(t)\Omega G \Omega S^T(s_2)H}_{j,i}&=\bb{S(s_1)\Omega Ge^{-\Omega Gt} \Omega S^T(s_2)H}_{j,i}\\
    &=\bb{S(s_1)\Omega G \Omega(e^{\Omega Gt})^{T} S^T(s_2)H}_{j,i}\\
    &=\bb{HS(s_2)e^{\Omega Gt}\Omega G \Omega S^T(s_1)}_{i,j}\\
    &=\bb{HS(s_2)T(-t)\Omega G \Omega S^T(s_1)}_{i,j},
\end{align}
where we used the fact that $H$ and $G$ are symmetric, and $\Omega^T=-\Omega$. With this,~\eqref{eq:result2} becomes
\begin{multline}
    [I_{\alpha, z}(r, H)]_{(j,k),i}=\frac{1}{2}\int_{-\infty}^\infty dt\int_0^1 ds_1\int_0^1 ds_2\,p_{\alpha,z}(t)\\
    \times \left(\bb{HS(s_2)T(-t)\Omega G \Omega S^T(s_1)}_{i,j}\bb{S(s_1)(\mu-r)}_k\right.\\
    +\left.\bb{HS(s_2)T(-t)\Omega G \Omega S^T(s_1)}_{i,k}\bb{S(s_1)(\mu-r)}_j\right).
\end{multline}
Because we are integrating $t$ over $(-\infty,\infty)$, and because $p_{\alpha,z}(t)=p_{\alpha,z}(-t)$, we can do the change of variable $t\ra-t$ and the integral remains unchanged. Thus, after interchanging the integration variables $s_1$ and $s_2$ and changing their order of integration, we find that
\begin{multline}
    [I_{\alpha, z}(r, H)]_{(j,k),i}=\frac{1}{2}\int_{-\infty}^\infty dt\int_0^1 ds_1\int_0^1 ds_2\,p_{\alpha,z}(t)\\
    \times \left(\bb{HS(s_1)T(t)\Omega G \Omega S^T(s_2)}_{i,j}\bb{S(s_2)(\mu-r)}_k\right.\\
    +\left.\bb{HS(s_1)T(t)\Omega G \Omega S^T(s_2)}_{i,k}\bb{S(s_2)(\mu-r)}_j\right),
\end{multline}
which matches~\eqref{eq:cross_term1}, as expected, thus completing the proof of~\eqref{eq:final_info_mtx_3}.

\section{Proof of Proposition~\ref{prop:sld}}
\label{app:sld}

We will begin by finding the derivative of the parametrized state $\sigma(\phi)$ with respect to an entry $\phi_j$ of the parameter vector $\phi$. We will use the shorthand $\pd_j\coloneq\frac{\pd}{\pd\phi_j}$. Duhamel's formula for the partial derivative of a matrix exponential will be helpful:
\begin{align}
    \pd_x e^{A(x)}&=\int_0^1 du\,e^{(1-u)A(x)}\bp{\pd_x A(x)}e^{uA(x)}.
\end{align}
First, we will find the derivatives of $e^{\pm i\hat{H}(\phi)}$. Consider the following:
\begin{align}
    \pd_je^{- i\hat{H}(\phi)}&=\int_0^1du\,e^{- i\hat{H}(\phi)(1-u)}\bp{\pd_j(- i\hat{H}(\phi))}e^{- i\hat{H}(\phi)u} \\
    &=-i\int_0^1du\,e^{- i\hat{H}(\phi)(1-u)}\bp{\pd_j(\hat{H}(\phi))}e^{- i\hat{H}(\phi)u}\\
    &=-i\int_0^1du\,e^{- i\hat{H}(\phi)u}\bp{\pd_j(\hat{H}(\phi))}e^{- i\hat{H}(\phi)(1-u)}\\
    &=-i\bp{\int_0^1du\,e^{- i\hat{H}(\phi)u}\bp{\pd_j(\hat{H}(\phi))}e^{i\hat{H}(\phi)u}}e^{-i\hat{H}(\phi)}\\
    &=-i\Psi^\dag_\phi\bp{\pd_j(\hat{H}(\phi))}e^{-i\hat{H}(\phi)},
\end{align}
where $\Psi_\phi$ is defined in~\eqref{def:psi_channel_time_evo}.
Similarly,
\begin{align}
    \pd_je^{ i\hat{H}(\phi)}&=\int_0^1du\,e^{ i\hat{H}(\phi)(1-u)}\bp{\pd_j( i\hat{H}(\phi))}e^{ i\hat{H}(\phi)u} \\
    &=ie^{i\hat{H}(\phi)}\int_0^1du\,e^{- i\hat{H}(\phi)u}\bp{\pd_j(\hat{H}(\phi))}e^{i\hat{H}(\phi)u}\\
    &=ie^{i\hat{H}(\phi)}\Psi^\dag_{\phi}\bp{\pd_j(\hat{H}(\phi))}.
\end{align}
It follows that
\begin{align}
    \pd_j\sigma(\phi)&=\pd_j\bp{e^{-i\hat{H}(\phi)}\rho e^{i\hat{H}(\phi)}}\\
    &=\bp{\pd_je^{-i\hat{H}(\phi)}}\rho e^{i\hat{H}(\phi)}+e^{-i\hat{H}(\phi)}\rho\bp{\pd_j e^{i\hat{H}(\phi)}}\\
    &=-i\Psi_\phi^\dag\bp{\pd_j(\hat{H}(\phi))}e^{-i\hat{H}(\phi)}\rho e^{i\hat{H}(\phi)}+ie^{-i\hat{H}(\phi)}\rho e^{i\hat{H}(\phi)}\Psi^\dag_{\phi}\bp{\pd_j(\hat{H}(\phi))}\\
    &=i\bb{\sigma(\phi),\Psi^\dag_{\phi}\bp{\pd_j(\hat{H}(\phi))}}.
\end{align}
Let $\sigma(\phi)=\sum_k\lambda_k\op{k}{k}$ be the spectral decomposition of $\sigma(\phi)$. Then,
\begin{align}
    \ip{k}{\pd_j\sigma(\phi)\middle|\ell}&=i\bp{\lambda_k \ip{k}{\Psi^\dag_{\phi}\bp{\pd_j(\hat{H}(\phi))}\middle|\ell}-\lambda_\ell \ip{k}{\Psi^\dag_{\phi}\bp{\pd_j(\hat{H}(\phi))}\middle|\ell}}\\
    &=i(\lambda_k-\lambda_\ell)\ip{k}{\Psi^\dag_{\phi}\bp{\pd_j(\hat{H}(\phi))}\middle|\ell}.
\end{align}
Using~\eqref{def:sld}, we find that the SLD operator is
\begin{align}
    L^{(j)}(\phi)&=\sum_{k,\ell}2i\frac{\lambda_k-\lambda_\ell}{\lambda_{k}+\lambda_{\ell}}\ip{k}{\Psi^\dag_{\phi}\bp{\pd_j(\hat{H}(\phi))}\middle|\ell}\op{k}{\ell}.
\end{align}
Let $\hat{G}_\phi=\sum_k\mu_k\op{k}{k}$ be the spectral decomposition of $\hat{G}_\phi\coloneq e^{-i\hat{H}(\phi)}\hat{G}e^{i\hat{H}(\phi)}$, so that $\hat{G}_\phi$ and $\sigma(\phi)$ share an eigenbasis, and $\lambda_k=\frac{e^{-\mu_k}}{Z}$. Then,
\begin{align}
    \frac{\lambda_k-\lambda_\ell}{\lambda_{k}+\lambda_{\ell}}&=\frac{e^{-\mu_k}-e^{-\mu_\ell}}{e^{-\mu_k}+e^{-\mu_\ell}}\\
    &=\frac{e^{\mu_\ell-\mu_k}-1}{e^{\mu_\ell-\mu_k}+1}\\
    &=\frac{\tanh\bp{\frac{\mu_\ell-\mu_k}{2}}}{\frac{\mu_\ell-\mu_k}{2}}\frac{\mu_\ell-\mu_k}{2}\\
    &=\int_{\RR}dt\,p(t)e^{-i(\mu_\ell-\mu_k)t}\frac{\mu_\ell-\mu_k}{2},
\end{align}
where we used the fact that $\frac{\tanh(\omega/2)}{\omega/2}$ is the Fourier transform of $p(t)$ given in~\eqref{def:high_peak_tent_pd}~\cite[Appendix B]{anshu_2021}. It follows that
\begin{align}
    L^{(j)}(\phi)&=\sum_{k,\ell}i\int_{\RR}dt\,p(t)e^{-i(\mu_\ell-\mu_k)t}(\mu_\ell-\mu_k)\ip{k}{\Psi^\dag_{\phi}\bp{\pd_j(\hat{H}(\phi))}\middle|\ell}\op{k}{\ell}\\
    &=i\sum_{k,\ell}\int_{\RR}dt\,p(t)\left(e^{i\mu_k t}\op{k}{k}\,\Psi^\dag_{\phi}\bp{\pd_j(\hat{H}(\phi))}e^{-i\mu_\ell t}\mu_\ell\op{\ell}{\ell}\right. \notag\\
    &\hspace{3cm}\left.-\mu_ke^{i\mu_kt}\op{k}{k}\,\Psi^\dag_{\phi}\bp{\pd_j(\hat{H}(\phi))}e^{-i\mu_\ell t}\op{\ell}{\ell}   \right)\\
    &=i\int_{\RR}dt\,p(t)e^{i\hat{G}_\phi t}\Psi^\dag_{\phi}\bp{\pd_j(\hat{H}(\phi))}e^{-i\hat{G}_\phi t}\hat{G}_\phi  \notag\\
    &\hspace{1cm}-i\hat{G}_\phi \int_{\RR}dt\,p(t)e^{i\hat{G}_\phi t}\Psi^\dag_{\phi}\bp{\pd_j(\hat{H}(\phi))}e^{-i\hat{G}_\phi t}\\
    &=i\int_{\RR}dt\,p(t)e^{-i\hat{G}_\phi t}\Psi^\dag_{\phi}\bp{\pd_j(\hat{H}(\phi))}e^{i\hat{G}_\phi t}\hat{G}_\phi  \notag\\
    &\hspace{1cm}-i\hat{G}_\phi \int_{\RR}dt\,p(t)e^{-i\hat{G}_\phi t}\Psi^\dag_{\phi}\bp{\pd_j(\hat{H}(\phi))}e^{i\hat{G}_\phi t}\label{eq:sld_p_even}\\
    &=i\Phi_{p,\hat{G}_\phi }\bp{\Psi^\dag_{\phi}\bp{\pd_j(\hat{H} (\phi))}}\hat{G}_\phi -i\hat{G}_\phi \Phi_{p,\hat{G}_\phi }\bp{\Psi^\dag_{\phi}\bp{\pd_j(\hat{H}(\phi))}}\\
    &=i\bb{\Phi_{p,\hat{G}_\phi }\bp{\Psi^\dag_{\phi}\bp{\pd_j(\hat{H}(\phi))}},\hat{G}_\phi },
\end{align}
where we used the fact that $p(t)=p(-t)$ to arrive at~\eqref{eq:sld_p_even}, thus concluding the proof of Prop.~\ref{prop:sld}.

\end{document}